\documentclass[conference,compsoc]{IEEEtran}
\pdfoutput=1
\ifCLASSOPTIONcompsoc
  \usepackage[nocompress]{cite}
\else
  \usepackage{cite}
\fi
\ifCLASSINFOpdf
\else
\fi
\usepackage{url}
\usepackage{xspace}
\usepackage{tikz}
\usepackage{amsmath}
\usepackage{subfig}
\usepackage{multirow}

\usepackage{filecontents}

\usepackage{threeparttable}

\usepackage{enumitem}
\usepackage{bm}
\begin{document}
\newcommand{\psca}{\textsc{PSCA}\xspace}
\newcommand{\pscas}{\textsc{PSCAs}\xspace}
\newcommand{\psc}{\textsc{PSC}\xspace}
\newcommand{\pscs}{\textsc{PSCs}\xspace}
\newcommand{\ali}[1]{\textcolor{orange}{Ali: #1}}
\newcommand{\pouya}[1]{\textcolor{red}{Pouya: #1}}
\newcommand{\fixme}[1]{\textcolor{red}{FIXME: #1}}
\newcommand{\meng}[1]{\textcolor{green}{Meng: #1}}
\newcommand{\ulysse}[1]{\textcolor{blue}{Ulysse: #1}}
\newcommand{\todo}[1]{\textcolor{yellow}{To Do: #1}}
%
\title{\Large \bf Exploring Power Side-Channel Challenges in Embedded Systems Security}



%
\author{\IEEEauthorblockN{Pouya Narimani, Meng Wang, Ulysse Planta, Ali Abbasi}
\IEEEauthorblockA{CISPA Helmholtz Center for Information Security}}



\maketitle

\begin{abstract}
Power side-channel (\psc) attacks are widely used against embedded microcontrollers, particularly in cryptographic applications, to extract sensitive information. However, expanding the applications of \psc attacks to a broader systems security context, especially in embedded systems, faces significant challenges. These challenges include the need for specialized hardware setups to manage high noise levels in real-world targets and unrealistic assumptions regarding the attacker’s knowledge and capabilities. This paper systematically analyzes these challenges and introduces a novel signal-processing method that addresses key limitations, enabling effective \psc attacks in real-world embedded systems without requiring hardware modifications. We validate the proposed approach through experiments on real-world black-box embedded devices, verifying its potential to expand its usage in various systems security applications beyond traditional cryptographic applications.
\end{abstract}


%
\IEEEpeerreviewmaketitle

\section{Introduction}
\label{sec:introduction}

Power side-channel (\psc) attacks are typically used to extract secrets from embedded Microcontroller Units (MCUs). These attacks~\cite{chari2003template,brier2004correlation,schindler2005stochastic,gierlichs2008mutual} exploit information leakage from power usage of the embedded devices. The leakage is primarily due to the physical characteristics of electronic components, exhibiting varying behaviors depending on the logical values they process. Although encryption algorithms were the initial target of side-channel attacks in embedded systems~\cite{kocher1999differential}, in recent years, researchers have broadened the application of the \psc to other system security domains. Notably, \psc has been used for \textbf{instruction disassembly}~\cite{eisenbarth2010building,park2018power,narimani2021side} as a primary focus. Also, there are some limited applications in code execution tracking~\cite{liu2016code}, offline \textbf{fuzzing}~\cite{sperl2019side}, and \textbf{intrusion detection}~\cite{han2022hiding,liang2021practical}.



When comparing \psc's traditional use in hardware security with its potential in broader security applications, we can observe three primary areas of difference: \textbf{methodology}, \textbf{assumptions}, and \textbf{evaluation}. 

Methodologically, research in hardware security often operates in the frequency domain, while existing work on system security applications mostly operates in the time domain. Operating in the frequency domain has significant advantages, as it allows for the filtering of noises from other hardware components operating at different frequencies. While the shift to using the frequency domain in hardware security was motivated by the requirement of advanced signal processing algorithms to defeat masking techniques designed to prevent \psc against cryptographic elements, such a transformation did not happen in system security applications, and as a result, it did not benefit from advancements in signal processing. Instead, the research in this system security often relies on time domain and pattern matching~\cite{liang2021practical} to find sequences of signals in different time windows, which is inefficient.

Additionally, the assumptions that are being made between hardware security and system security are different. In the context of hardware security, it is generally presumed that the attacker is aware of the cryptographic algorithm’s implementation and has access to the target device or a similar one to study it before launching the attack. While these prerequisites might be acceptable in certain system security applications (such as designing our PCB for intrusion detection), they may not be suitable for other applications (e.g., fuzzing a third-party board). Therefore, the profiling-based methods face challenges in generalizing across different PCBs, especially for black-box embedded systems (i.e., embedded systems in which their firmware, source code, or Instruction Set Architecture (ISA) is inaccessible).


Finally, evaluation techniques in hardware security-focused \psc research differ significantly from those in the system security domain. Evaluations in hardware security are often conducted under highly controlled conditions, frequently using custom-built boards designed to minimize hardware noise~\cite{hori2012sasebo}. Such setups allow researchers to test the effectiveness of security measures like algorithmic masking techniques~\cite{dubey2020maskednet}. However, these controlled environments do not accurately reflect real-world conditions in system security, where noise and variability are substantial factors.

Interestingly, some system security applications of \psc adopt similarly controlled approaches for their evaluation. For instance, evaluations are often carried out on isolated MCUs rather than on integrated PCBs~\cite{liu2016code,narimani2021side}, sometimes using low-end microcontrollers~\cite{msgna2014precise,jungmin2019leveraging} (e.g., 8-bit or 16-bit) or boards modified to further reduce noise, such as through the addition of shunt resistors~\cite{han2022hiding}. In these setups, peripheral interactions are minimized or do not exist to limit hardware noise even further.
In contrast, power traces in real-world targets are considerably noisier, incorporating interference from various hardware (peripheral interactions, hardware components on the board) and firmware components (e.g. context switches) and running on more complex MCUs.

Based on the above observations, we argue that most research in extending the application of \psc beyond traditional cryptographic applications contains unrealistic assumptions, methodologies, and evaluation.
Inspired by the work of Muench et al.~\cite{muench2018you} on addressing challenges in fuzzing embedded systems firmware, our objective in this work is twofold:

First, we identify the existing challenges that lead to impractical methodologies, assumptions, and evaluations in \psc applications within the system security context~\cite{sperl2019side, han2022hiding, narimani2021side}. These challenges may be broad (such as the pervasive issue of noise), or they may pertain to specific applications, like real-time requirements in fuzzing. Furthermore, we explore the application of \pscs in system security. As a result, we systematically analyze existing \psc-based attacker models within the embedded systems domain and explore why current research struggles to generalize across different contexts. 

Second, we propose a novel signal-processing algorithm specially designed for system security applications to process the power traces in the frequency domain and suppress noise both in hardware and software. Notably, our method is the \textit{first} to work on real-world black-box embedded devices without requiring any hardware modifications (e.g., shunt resistor or high-resolution ADC) for side-channel probing. Additionally, our method generalizes across various board designs without the need for target profiling or performing template attacks, something that does not exist today.

To the best of our knowledge, we are the first to apply power side-channel on high-end embedded systems (an application processor CPU clocked at 1Ghz with a Cortex-A 64bit ARM CPU instead of traditional Microcontrollers). As a result of this novel algorithm, we explore additional improvements to existing system security applications, namely fuzzing and intrusion detection, and introduce another application, namely software composition analysis, which allows us to detect algorithms running inside real-world embedded devices without access to the firmware or performing template attack to model the MCU. We will conduct thorough evaluations to demonstrate these results and underscore the potential for further extending the application of \psc across additional system security domains.

To foster research on this topic, we release the experimental data sets and our signal processing algorithm implementation on Github~\cite{paper-repo}.
In summary, we make the following contributions:
\begin{itemize}[noitemsep]
    \item
    We perform a first comprehensive and systematic analysis of the existing \psc applications for embedded devices system security.
    \item
    We identify ten key attacker models and five applications that exploit \psc in embedded devices.
    \item
    We discuss and demonstrate the general limitations of \psc and analyze their effectiveness in the software security domain through experiments.
    \item
    We propose a novel solution based on signal processing to address existing challenges and evaluate them on real-world targets, including high-end processors.
\end{itemize}

\section{Background}
\label{sec:background}

\begin{figure*}[!h]
    \centering
    \subfloat[\centering Raw Power Trace in Time Domain]{{\includegraphics[width=4.4cm]{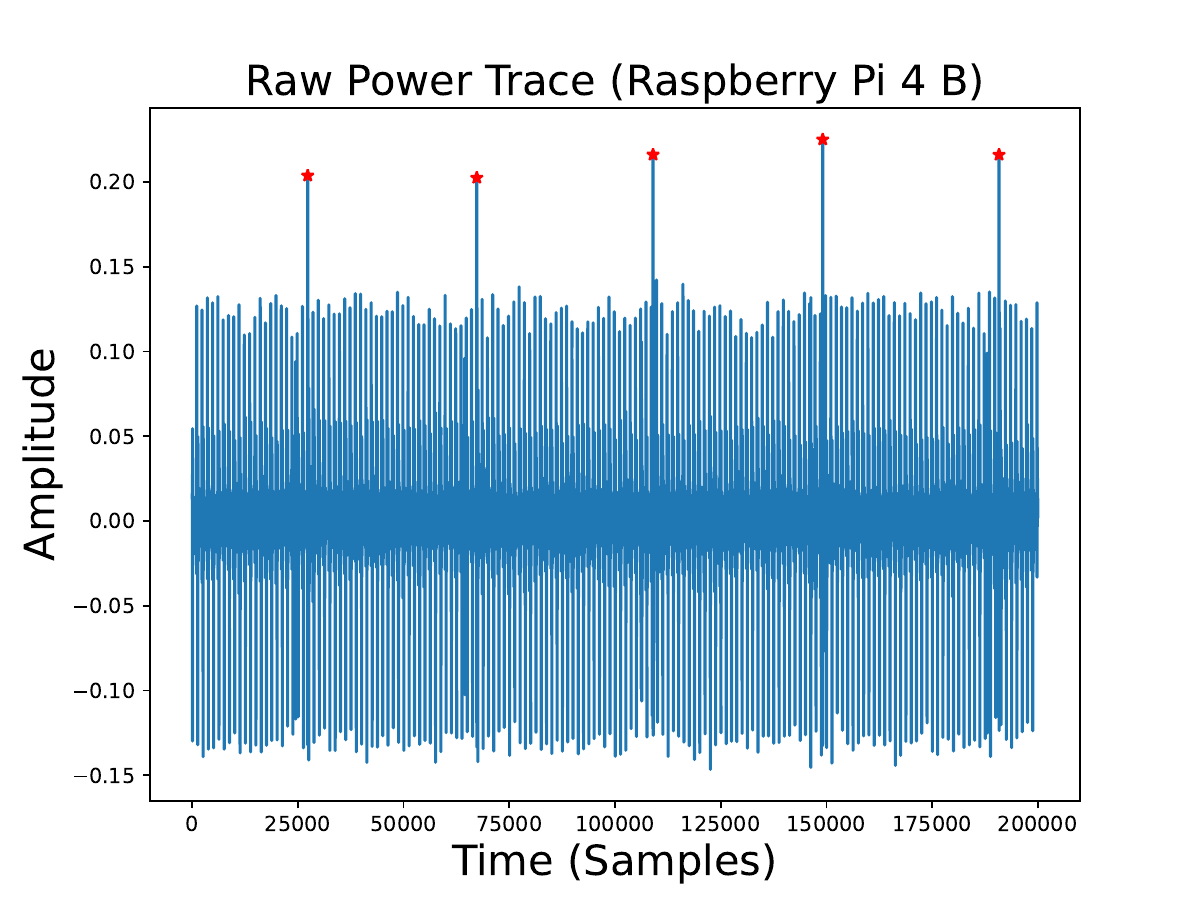}}}
    \subfloat[\centering FFT and Bandpass Filter]{{\includegraphics[width=4.4cm]{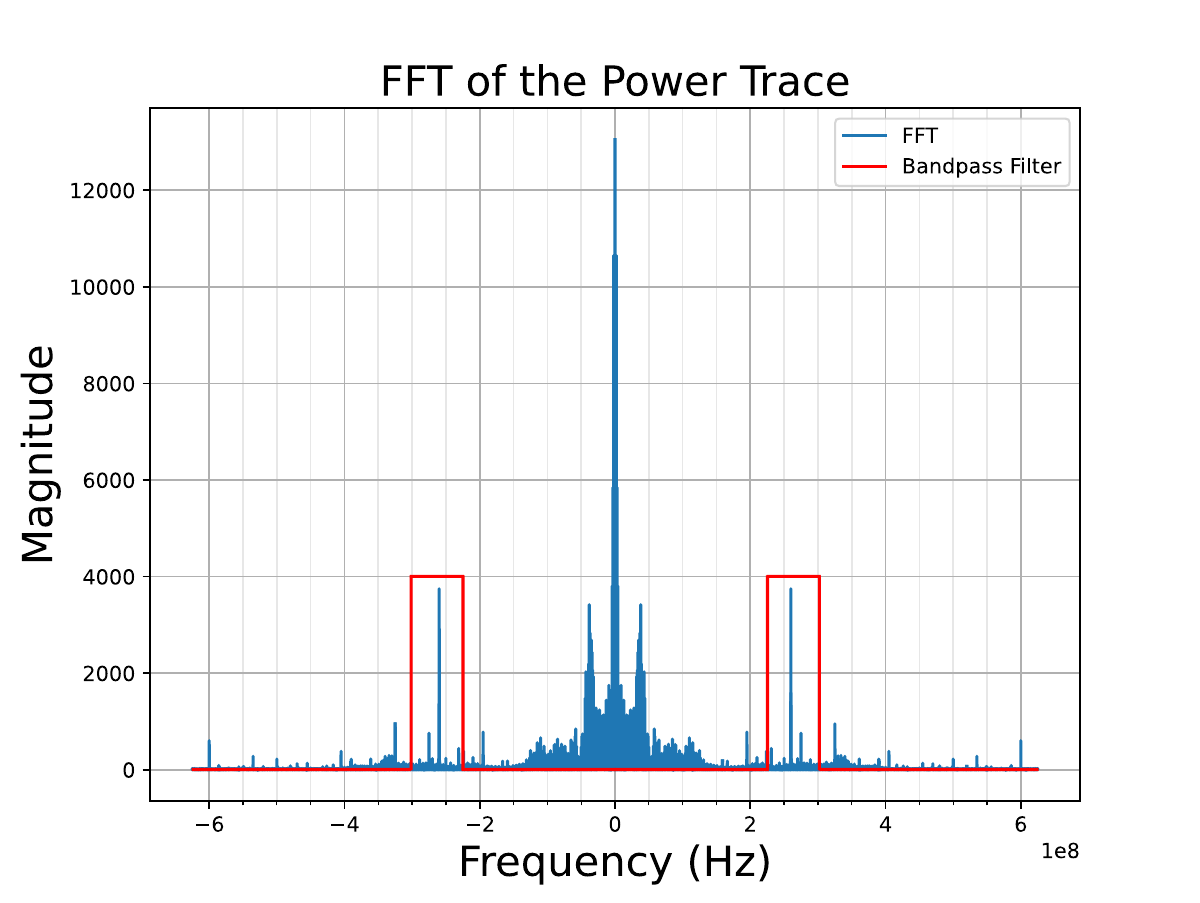}}}
    \subfloat[\centering Filtered Power Trace]{{\includegraphics[width=4.4cm]{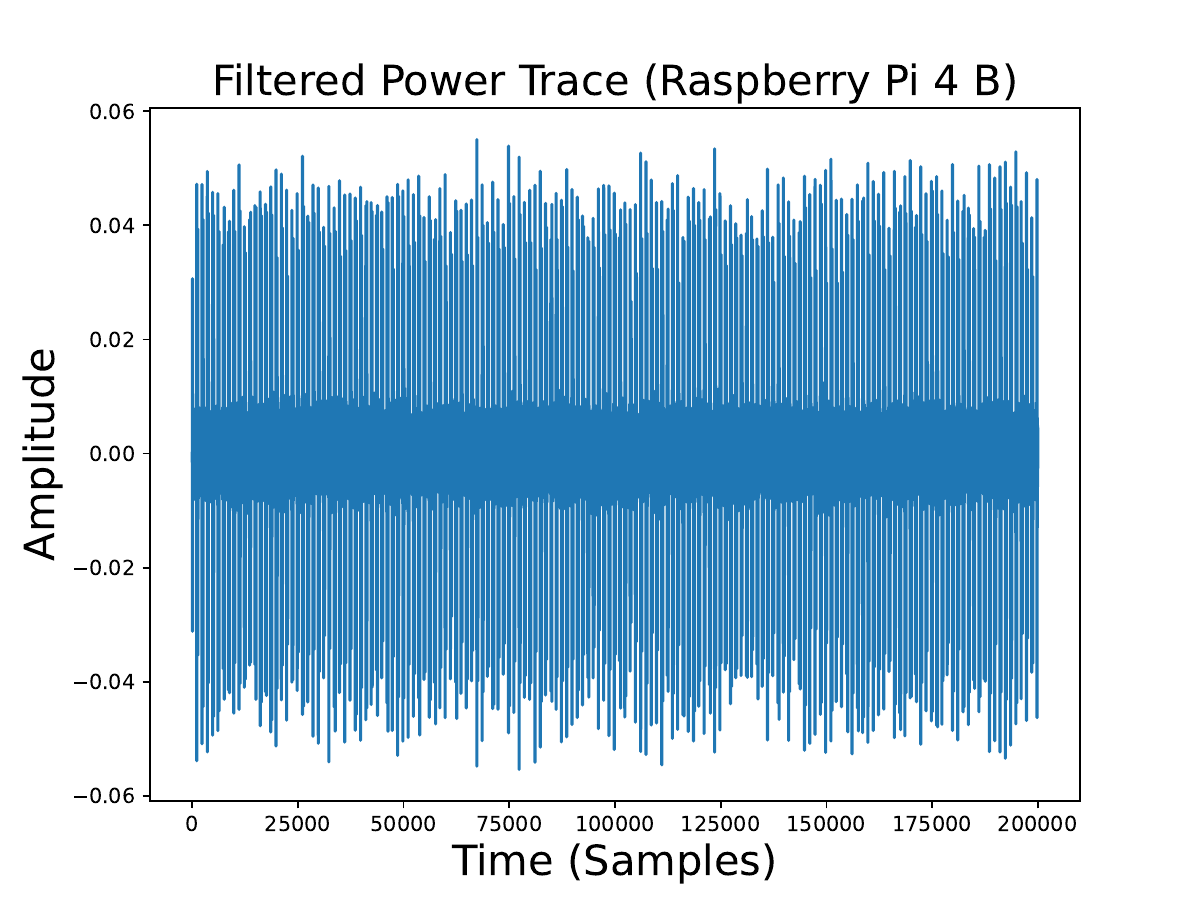}}}
    
    \caption{Frequency domain representation shows a better view of the signal that can help distinguish power leakage originating from different sources of the target board.}
    \label{fig:Freq_Domain}
\end{figure*}

This section outlines the key concepts that will be used in the following sections, including the concept of \psc profiling and frequency domain representation and transformations.
\subsection{\psc Profiling}
\label{sec:background:profiling}
\psc leak information as a result of signal-switching within integrated circuits. The extent of data that can be extracted from power traces depends on both the magnitude of the data's effect on the power trace and the Signal-to-Noise Ratio (SNR) of the captured signal. Depending on the attacker's capabilities and access, different types of \psc attacks can get implemented. These types may vary in terms of their complexity, the level of knowledge the attacker has about the target, and the methods used to exploit the leakage.


One key distinction in side-channel attacks is between profiling and non-profiling attacks. In a profiling attack, the attacker has access to a replica of the target device, which they can use to develop a leakage model~\cite{chari2003template,schindler2005stochastic}. This leakage model is then applied to the actual target during the attack phase. Profiled side-channel attacks are considered the most powerful type of side-channel attack, as they assume a worst-case scenario where the attacker has unlimited access to the target device during the profiling phase.




\subsection{Frequency Domain Transformations}
\label{sec:background:frequency:domain}
\psc traces are time series that capture the changes in voltage over time during the execution of a process. The \textit{time domain} representation of these traces shows how the signal evolves as time progresses. This view helps better pattern visualization and signal analysis. However, analyzing signals purely in the time domain can be computationally expensive and may not reveal important characteristics of the signal.
To make signal analysis more efficient and informative, we can transform the signal from the time domain to the \textit{frequency domain}. This transformation is commonly done using methods like the Fourier Transform~\cite{bracewell1989fourier}, which breaks the signal down into its component frequencies. In other words, it shows which frequencies are present in the signal and how much each frequency contributes to the overall signal. This frequency-based representation provides a different, often more intuitive, view of the signal, allowing for better signal processing and analysis by highlighting patterns that are not as obvious in the time domain.
One significant advantage of working in the frequency domain is that certain operations are simplified in this domain (for example, convolution in the time domain is transformed into a multiplication in the frequency domain). As an example, we recorded the power trace of a Raspberry Pi 4B while writing to memory. As we show in Figure~\ref{fig:Freq_Domain} (a), the power trace has massive peaks (marked with red stars) that disturb the signal, while the smaller peaks in between are due to memory interaction. However, by transforming the signal to the frequency domain (Figure~\ref{fig:Freq_Domain} (b)), we distinguish those peaks and thus filter them using a bandpass filter. Then, we return the filtered signal to the time domain (Figure~\ref{fig:Freq_Domain} (c)), which shows a clean signal with only memory interactions.
Furthermore, various signal-processing techniques are specifically designed for the frequency domain. For instance, the wavelet transform~\cite{daubechies1990wavelet} is a well-known method used to represent a signal in terms of both the existing frequencies and a "wavelet" function. In simple terms, the wavelet transform scans the time series, looking for repeating patterns that match the base wavelet function. This makes the wavelet transform particularly powerful because it analyzes the frequencies in a signal and retains some information about the original time series. Unlike other methods, such as the Fourier transform, which focuses purely on frequencies and loses the connection to when those frequencies occur, the wavelet transform allows you to see both the frequency content (the different frequencies that make up the signal) and wherein the time series those frequencies are present. 

\subsection{Low-end and High-end Embedded Devices}

Embedded devices are divided into two categories: low-end and high-end. Low-end devices typically have a microcontroller unit (MCU) and a single core with RAM integrated inside the chip. On the other hand, high-end devices refer to those with a microprocessor unit (MPU). MPUs usually have multiple complex cores with cache and external RAM. They operate at higher frequencies compared to MCUs and have more complex architectures. Existing work in \psc for system security focuses on low-end targets, while nowadays high-end embedded devices are widespread.






\section{General Limitations}
\label{sec:general:limitations}
To understand the general limitations of \psc attacks on embedded devices, we start by examining the common challenges and reviewing the solutions proposed by the research community. Some limitations are observable across various applications. We evaluate these application-independent limitations and discuss how they affect the effectiveness of \psc attacks on embedded systems.

\subsection{Probing}
\label{sec:general:limitations:probing}
PSCs are usually measured using a shunt resistor (1 to 100 Ohm, depending on the current consumption of the processor) on GND (ground) or main power supply (VCC or VDD) pins.
A shunt resistor is a resistor placed on the power supply lines of the target processor, causing a small voltage drop between the main power supply lines and the processor. The voltage drop caused by the processor's power consumption, which is different for various bit sequences, is the main source of \psc leakage. The ideal probing point for \psc leakage is after the shunt resistor.
However, real-world circuit boards targeted by \psc attacks usually don't have any shunt resistor. Therefore, the attacker needs to either add a shunt resistor on the processor pins by manipulating the target board or measure the power traces on the main power supply of the board by adding a shunt resistor on it. In the first case, power traces are less noisy as the voltage drop does not have any effects from other components, while in the second one, the noise of other components on the Printed Circuit Board (PCB) affects the \psc traces. Hence, the probing point is very important as it significantly affects the information leakage. Consequently, as the probing point gets closer to the target processor by bypassing other components on the PCB, the amount of noise will decrease, and the measurements will have more information. Therefore, choosing a proper point for \psc probing directly affects the accuracy of an attack. As a result, to facilitate accurate measurements, some hardware platforms have been presented. For instance, ChipWhisperer (CW)~\cite{ChipWhispererHusky} is a board specifically designed for \psc attacks, featuring a soldered shunt resistor for precise measurements. 

\subsection{Noise}
\label{sec:general:limitations:noise}
\begin{table*}[!ht]
    \captionsetup{font=footnotesize}
    \centering
    \footnotesize
    \begin{threeparttable}

    \begin{tabular}{|c|c|c|c|c|c|c|} 
\hline
\multicolumn{1}{|c|}{Application}              & ISA & Firmware & Semantics of Interface & Offensive & Category & (Non) Profiling \\ 
\hline
\multirow{4}{*}{IDS}                           & \multicolumn{1}{c|}{Unknown} & \multicolumn{1}{c|}{N/A} & \multicolumn{1}{c|}{Unknown} & \multicolumn{1}{c|}{No}  & \multicolumn{1}{c|}{BB}  & \multicolumn{1}{c|}{Profiling} \\ 
\cline{2-7}
& Unknown & N/A & Known & No & GB1 & Profiling \\ 
\cline{2-7}                                    & \multicolumn{1}{c|}{Known} & \multicolumn{1}{c|}{Available} & \multicolumn{1}{c|}{Unknown}   & \multicolumn{1}{c|}{No}  & \multicolumn{1}{c|}{GB2} & \multicolumn{1}{c|}{Profiling}      \\ 
\cline{2-7}
 & Known & Available & Known & No & WB & Profiling \\ 
\hline
\multicolumn{1}{|c|}{\multirow{3}{*}{Fuzzing}} & Known & Available & Known & Yes & WB & Non Profiling \\ 
\cline{2-7}
\multicolumn{1}{|c|}{}                         & Unknown & N/A & Known & Yes & GB & Non Profiling \\ 
\cline{2-7}
\multicolumn{1}{|c|}{}                         & Unknown & N/A & Unknown & Yes & BB & Non Profiling \\ 
\hline
\multicolumn{1}{|c|}{Disassembly}              & Known & N/A & Unknown & Yes & GB & Profiling \\ 
\hline
SCA\tnote{1}                                   & \multicolumn{1}{c|}{(Un)known} & \multicolumn{1}{c|}{N/A}    & \multicolumn{1}{c|}{(Un)known} & \multicolumn{1}{c|}{Yes} & \multicolumn{1}{c|}{BB}  & \multicolumn{1}{c|}{Profiling}     \\ 
\hline
Peripheral Interaction                         & \multicolumn{1}{c|}{(Un)known} & \multicolumn{1}{c|}{N/A}    & \multicolumn{1}{c|}{(Un)known} & \multicolumn{1}{c|}{Yes} & \multicolumn{1}{c|}{BB}  & \multicolumn{1}{c|}{Non Profiling}  \\
\hline
\end{tabular}
\begin{tablenotes}
    \item[1] Software Composition Analysis
\end{tablenotes}
    \caption{This table outlines attacker models across applications, categorized by the attacker capabilities. Each row represents a threat model and is categorized as white-box (WB), gray-box (GB), or black-box (BB). For example, a gray-box attacker fuzzing an embedded system is defined as a non-profiling attacker with only knowledge of how to interface with the device, lacking any knowledge of the ISA and the firmware.}
    \label{tab:attacker_model}
    \end{threeparttable}
\end{table*}

Electrical noise is an inherent aspect of all electrical circuits and cannot be eliminated, but its impact can be minimized depending on the application. In the context of \psc attacks, electrical noise is a crucial factor influencing scalability. The noise level in the \psc can vary based on the characteristics and components of the circuit board. Consequently, generalizing results to other boards with different designs or components, even if they use the same target processor, is often challenging or impossible in many \psc attacks. Therefore, \psc attacks are highly dependent on the specific design and characteristics of the target board.
Furthermore, filtering noise from useful information is particularly challenging because \psc leakages often occur at noise levels, making it difficult to distinguish between the two. This means that attackers must not only analyze the impact of noise on the target device but also conduct extra measurements (depending on the complexity of the target and the resolution of the measurement setup) to mitigate the noise effect. In the \psc, noise can originate from various sources. Besides the noise from components, interactions with peripherals can also lead to abnormal power consumption, which can affect measurements in different ways depending on the application. Therefore, managing each type of noise requires tailored strategies to reduce its impact.

\subsection{Signal Processing}
\label{sec:general:limitations:signal:processing}

As highlighted in two previous limitations, the probing point and the resulting noise level significantly influence the amount of information contained in \psc traces. A common approach to mitigating noise is to repeat measurements and average the results to reduce noise. For instance, when targeting a cryptographic algorithm, an attacker might take multiple measurements with different inputs, all encrypted with the same key. Averaging these \psc traces helps to reduce the noise.


The required number of measurements is influenced by various factors, including the probing point, the complexity of the target PCB, the resolution of the power traces, and the architectural complexity of the processor, such as the number of pipeline stages. Although the community has proposed several signal processing and statistical techniques~\cite{hajra2014multivariate, das2019x, picek2017side, bruneau2015less} to reduce the number of measurements and enhance the SNR, it remains challenging to execute an attack using a single power trace in most applications.


Additionally, power trace alignment presents another challenge in attacking cryptographic algorithms. Alignment ensures that power traces are horizontally synchronized so that the execution points of intermediate variables occur at the same time. Introducing misalignment into power traces is a countermeasure against \psc attacks, typically achieved by adding random dummy operations or shuffling operations~\cite{guneysu2011generic} to decrease the likelihood of a successful attack. There has been extensive research addressing these limitations~\cite{hettwer2020applications, cagli2017convolutional}. However, this paper will not focus on these solutions as they fall outside its scope.


\section{Challenges in Embedded Systems Security}
\label{sec:challenges:sysec}

The primary objective of this paper is to examine the application of \psc attacks in the context of embedded systems security. Having discussed the general limitations, we will now explore how these limitations impact the system security domain and analyze their effects (Addressing \textbf{RQ2}). To achieve this, we will begin by categorizing the existing applications of \psc attacks within embedded system security.


\subsection{Attacker Model}
\label{sec:challenges:sysec:attacker:model}


In the embedded system security domain, \pscs can be applied to various use cases, each with distinct attacker models. Table~\ref{tab:attacker_model} outlines the conditions under which \pscs can be effectively employed, along with the corresponding attacker models. We identify three key conditions that determine the applicability of \pscs for specific purposes, namely:
\begin{itemize}[noitemsep]
    \item {ISA:}
    Whether the ISA is known or unknown to the attacker.
    
    \item {Firmware:}
    Whether the firmware image is available to the attacker or not.
    \item {Semantics of Interface:}
    Whether the semantics of the communication interface of the target is known or unknown to the attacker.
\end{itemize}
Additionally, each attack is categorized as white-box, gray-box, or black-box, indicating the level of knowledge and access the attacker has in the attack scenario. In this context, we exclude scenarios where the firmware's source code is available, as rehosting techniques are more effective for fuzzing in such cases. Similarly, for Intrusion Detection Systems (IDS) and disassembly use cases, if the source code is accessible, the need for \psc analysis is eliminated.


The attacker models in existing research are not generalizable to real-world applications as, in most real-world cases, the firmware or ISA or both are not available. However, none of the existing research evaluates their method in real-world scenarios, leaving the efficiency of their method in such scenarios unknown. As an example, considering a Siemens Programmable Logic Controller (PLC) as a target, whose ISA and firmware are unknown, none of the existing \psc-based disassemblers can disassemble its firmware.


Regarding the mentioned defects in existing attacker models, our attacker model in this paper has the following assumptions:
\begin{itemize}[noitemsep]
    \item
    The attacker does not know the firmware or ISA of the target, he knows only the semantics of the interface.
    \item
    The attacker does not have to profile the same device.
    \item
    The attack is non-invasive, meaning that the attacker does not manipulate the target (no shunt resistor).
    \item
    The targets are real-world, off-the-shelf embedded devices without any custom design for \psc measurement.
\end{itemize}

\subsection{\psc Applications}
\label{sec:challenges:sysec:applications}

We can envision various applications of \psc in embedded system security. Some of these applications are introduced in the literature, while others are suggested in this paper. This section addresses \textbf{RQ1}.

\subsubsection{Disassembly:}
\begin{table*}[!ht]
\captionsetup{font=footnotesize}
    \centering
    \footnotesize
    \begin{tabular}{|c|c|c|c|c|c|c|c|c|}
         \hline
          Research & Method & Hardware & Bus Width & Dataset & Profiled Instructions & Cross-Device & Accuracy (\%)\\
          \hline
          Msgna et~al.~\cite{msgna2014precise} & KNN & ATMega163 & 8-bit & Random & 39/130 & No & 100 \\
          \hline
          Jungmin et~al.~\cite{jungmin2019leveraging} & SVM/QDA & ATMega328p & 8-bit & Random & 112/131 & No & 99.03 \\
          \hline
          Eisenbarth et~al.~\cite{eisenbarth2010building} & LDA & PIC & 8-bit & Random & - & No & 58 \\
          \hline
          Geest et~al.~\cite{van2022side} & CNN/MLP & ARM-M0 & 32-bit & Random & 17/57 & No & 99 \\
          \hline
          Narimani et~al.~\cite{narimani2207novel} & CNN & AVR & 8-bit & Random & 109/130 & No & 54.58 \\
         \hline
    \end{tabular}
    \caption{Existing \psc-based disassemblers in the literature are reported in this table.}
    \label{tab:disassembly_compare}
\end{table*}
Disassembly attacks focus on executed instruction sequences within firmware, aiming to recover its contents. These attacks can be classified into two types: \textit{instruction-level} and \textit{bit-level}

\textit{Instruction-level} attacks involve modeling the power consumption pattern of individual instructions to create a reference template for \psc patterns. During the attack phase, the attacker leverages these templates to identify the executed instructions~\cite{eisenbarth2010building}. ML models outperform other methods in this context, as they often generalize more effectively than traditional template-based approaches~\cite{park2018power, narimani2021side}.
\textit{bit-level} approaches, by contrast, focus on extracting bits of state from the processor pipeline. Each instruction consists of a sequence of bits that includes the opcode and operand(s); detecting these bits can potentially reveal the entire instruction~\cite{cristiani2020bit}. 
Table~\ref{tab:disassembly_compare} provides an overview of existing \psc-based disassemblers, comparing their methodologies, target hardware, datasets, the number of instructions targeted from the ISA, cross-device evaluation, and accuracy. A key limitation of current disassemblers is their focus on simple processors, usually with just two pipeline stages, and their lack of testing on applications with a Real-Time Operating System (RTOS), which is common in real-world embedded systems. Consequently, their effectiveness in real-world scenarios remains largely unassessed. As outlined in Table~\ref{tab:attacker_model}, the attacker model for these studies generally assumes a gray-box approach, where the attacker has knowledge of the ISA and firmware during profiling. However, this model does not extend well to black-box scenarios, where such information is unknown to the attacker.
    
\subsubsection{Control Flow Graph:}
As mentioned in the previous subsection, the \psc leaks information about the instructions being executed in the program. As a result, one can use this information to construct the Control Flow Graph (CFG) of the application and monitor its different run-time behaviors. For example, CFG data can be used to recognize abnormal patterns to detect intrusion (i.e., IDS)~\cite{han2017watch, han2022hiding},
or it can be used to check the program's integrity~\cite{liu2016code}.
Here, the implemented model should be lightweight enough to operate in real-time, especially in critical industrial applications~\cite{liang2021practical}.
The attacker model of the mentioned intrusion detection systems are white-box as they have access to the ISA and firmware of the target.
\subsubsection{Fuzzing:}
Fuzzing has been recognized as one of the most efficient and practical methods to find security vulnerabilities in both hardware and software. Considering that the \psc leaks information about the CFG, it can be used to guide a fuzzer. Recently, \psc has been used as a coverage feedback mechanism to guide a fuzzer in a theoretical way~\cite{sperl2019side}. The attacker model in this work is a white-box model as they have access to the ISA and firmware in profiling phase, and they use the same setting in the attack phase. However, this attacker model does not generalize to real-world scenarios as usually the attacker does not have ISA and/or firmware.
Detecting conditional branch instructions is crucial in coverage-guided fuzzing because they often signal the possibility of new basic blocks and new execution paths. This information collected from \psc can guide a fuzzer in generating inputs that explore a broader portion of the program, helping to identify possible security vulnerabilities across the program’s entire space.
Meanwhile, crash detection and root cause analysis in embedded devices is one of the open problems in embedded fuzzing. The existing challenges regarding embedded fuzzing have already been discussed in~\cite{muench2018you}, therefore we skip it here.
\subsubsection{Software Composition Analysis:}
\psc can enable software composition analysis on embedded devices without requiring firmware access. By analyzing power consumption patterns during execution, it may be possible to identify the presence and behavior of known third-party software components. Since different software modules exhibit distinct power usage profiles, these patterns can help determine which libraries or components are running.

\subsubsection{Peripheral Interaction:}
Another application of the PSC in embedded devices is detecting peripheral interactions. Generally, a microprocessor communicates with external devices using its various communication interfaces, such as UART, SPI, and Ethernet. These devices can be either on the PCB, such as an SPI flash, or they can be external devices with a separate PCB, such as a slave processor or sensor. In general, the power consumption for these communications is significantly more than normal internal processes as these interfaces communicate with external devices. Therefore, if the attacker monitors the power consumption, then they can decode the data that is being transferred using these interfaces. In the case that the external device is on the same PCB (the SPI flash example), the pins might not be accessible for decoding the data lines due to the BGA packaging. While the attacker can directly monitor the power pins from different locations and decode the data.

\subsection{Experimental Setup}
\label{sec:challenges:sysec:expsetup}

We design an experiment for each challenge we introduce to demonstrate it in the context of embedded system security. We use a Tektronix MSO66B oscilloscope equipped with passive probes that offer a 1 GHz bandwidth (Appendix~\ref{sec:appendix:exp_setup}). The sampling frequency varies depending on the experiment and satisfying Nyquist criterion, ranging from 25 MS/s to 3.125 GS/s.

For our experiments, we select five different targets to cover different scenarios in \psc measurements:
\begin{itemize}[noitemsep]
\item \textbf{Ideal Target}: A STM32F3 board with a shunt resistor soldered directly onto the PCB and without any peripherals, providing an ideal setup for precise \psc measurements.
\item \textbf{General-Purpose Development Board}: An ARM Cortex-M4 board with a STM32F401 microcontroller without soldered shunt resistors with various peripherals, representing a more realistic target.
\item \textbf{General-Purpose Development Board (Nucleo)}: A general-purpose development board, Nucleo-144~\cite{Nucleo144}, without any \psc measurement considerations, such as a shunt resistor. This board features a Cortex-M4 STM32F429 microcontroller, an onboard ST-LINK programmer and debugger, and an Ethernet socket for network communications.
\item \textbf{Real-World Black-Box Target (UBlox)}: A UBlox ZED-F9P-02B GNSS receiver~\cite{ubloxmodule} is used as a real-world black-box embedded target. This device's firmware and ISA are unavailable, and only the interface semantics (protocol specifications) are available, making it a suitable target for testing \psc analysis in scenarios with limited information.
\item \textbf{Real-World Black-Box Target (Satellite)}: We employ a satellite on-board RF-transceiver to evaluate the application of \psc for observing inaccessible data busses (e.g., busses on internal layers of a multi-layer PCB) non-destructively. Other than what can be identified from the markings of chips on the PCB, the circuits and the device's firmware are entirely unknown.
\item \textbf{Real-World High-end Target (Raspberry Pi)}: We use a Raspberry Pi Zero 2 W~\cite{RasPi} board as a high-end target (with a 64-bit ARM Cortex-A53 processor clocked at 1GHz) running a raspbian kernel. This board has a processor power test pin, which we use for power measurement.
\end{itemize}

\subsection{Probing}
\label{sec:challenges:sysec:probing}

In the embedded system security domain, selecting a probing point is significantly more challenging than in other domains. For example, in cryptographic attacks, the attacker uses an ideal board designed for \psc measurements (such as \cite{hori2012sasebo, katashita2012side}). Similarly, in defensive scenarios such as IDS, the defender could enable the inclusion of optimal probing points for more reliable \psc analysis in the PCB design stage. However, if the defender wants to deploy the \psc-based IDS to an existing embedded device, finding an ideal probing point would requires hardware manipulation.
Typically, attackers are dealing with an existing embedded device where they aim to deploy an attack. In these cases, provisions such as a shunt resistor for \psc measurements are absent, forcing the attacker to either add these requirements (often impractical) or resort to a noisy probing point on the device's main power supply.

\paragraph{\textbf{Experiment}}
To show the impact of noise on the probing point, we design an experiment. We choose two targets: the first one is the ideal target, and the second one is the STM32F4 development board with a shunt resistor manually attached to the main power supply lines of the board. We run a program with random combinations of four instructions—\texttt{ADD}, \texttt{SUB}, \texttt{AND}, and \texttt{ORR}—on both boards and record the power traces. The sampling frequency is set to 1.25 GS/s, and the bandwidth is 20 MHz. For classification, we use a Multi-Layer Perceptron (MLP), a method previously employed for \psc disassembly in \cite{van2022side}. On the target with the soldered shunt resistor, the classification accuracy is 93\%, while on the development board, the accuracy drops to 51\% (Figure~\ref{fig:DirtyClean_Noise} in Appendix~\ref{sec:appendix:CM1}). This result demonstrates that the presence of other electrical components on the PCB introduces significant noise, which drastically reduces the accuracy of instruction disassembly when using the same method.

\subsection{Noise}
\label{sec:challenges:syssec:noise}

\begin{figure}[!t]

    \centering
    \subfloat[\centering UBlox Power Trace]{{\includegraphics[width=4.4cm]{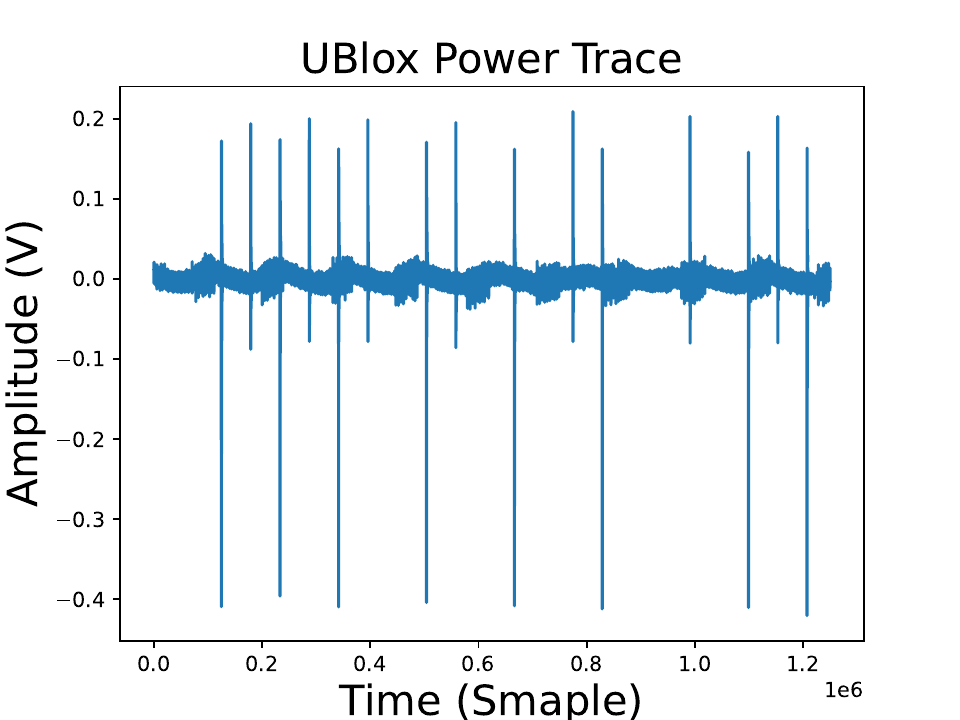}}}
    \subfloat[\centering Zoomed view in on one of the peaks]{{\includegraphics[width=4.4cm]{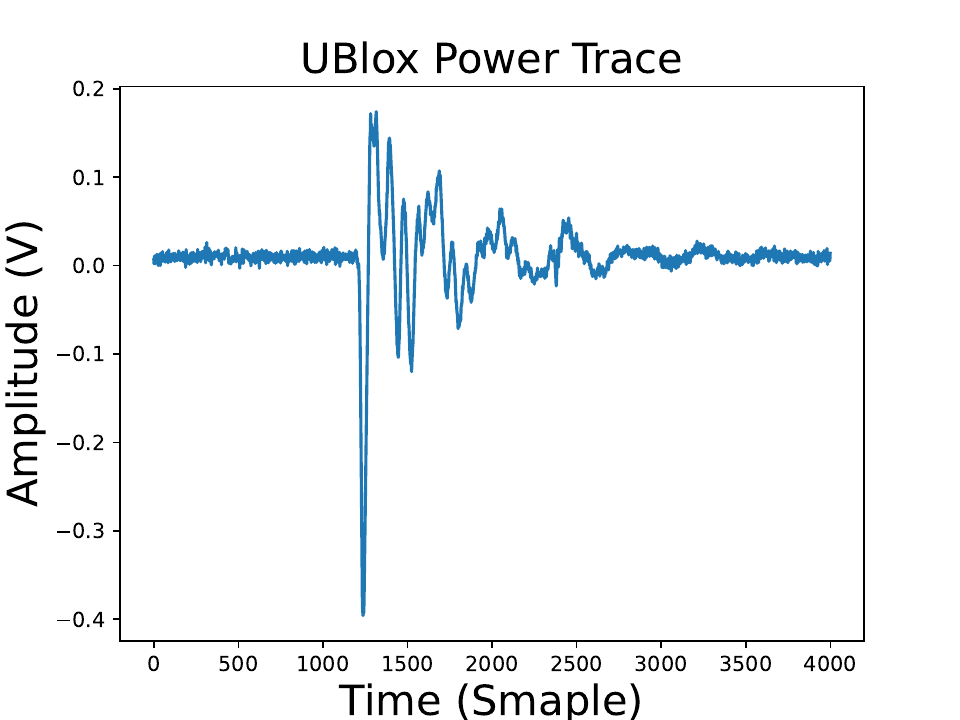}}}
    \caption{Peripheral interaction's effect in power trace}
    \label{fig:PeripheralInteraction}
\end{figure}

As discussed before, noise is unavoidable in \psc measurements. Here, noise includes any type of disturbance in the power signal that affects the power trace patterns and reduces its information. In an embedded device, the board has multiple peripherals and components on it. The interaction of peripherals, such as serial communication or GPIO-related functions, is detectable in power traces. This leads to problems in various applications, such as instruction disassembly or fuzzing, as it changes the power patterns. Thus, applying traditional signal processing methods does not work well when we have these disturbances. For instance, the proposed IDS in~\cite{han2022hiding} is only evaluated on the CW which does not have any peripherals. Consequently, power traces do not have any peripheral-related noise. Hence, the performance of the proposed method's accuracy can decrease in real-world targets with peripheral interactions.

\paragraph{\textbf{Experiment}}
To illustrate this issue, we choose the UBlox GNSS receiver as a real-world embedded device. This board has multiple serial interfaces for communication. We put a shunt resistor on the main power supply lines of the board and set the sampling frequency to 1.25 GS/s. Figure~\ref{fig:PeripheralInteraction} shows the recorded power traces during the UART communication. The high peaks demonstrate the state changes in the UART signal, which disturbs the main power trace.
Additionally, the harmonics of these peaks affect multiple execution cycles afterward. This raises problems in applications such as instruction disassembly and fuzzing as the power pattern for those instructions is disturbed. This effect is only desired in peripheral interaction, which can be used to decode the data.


\subsection{Signal Processing}
\label{sec:challenges:syssec:signalprocessing}

Traditional \psc attacks use various signal processing techniques to enhance the Signal-to-Noise Ration (SNR) and improve attack efficiency. However, in the context of embedded firmware, which often includes an embedded RTOS, additional background processes introduce extra noise that complicates the signal processing. This noise is primarily timing-related, causing misalignment in the power traces. It is important to note that this misalignment is unintentional and differs from the intentional misalignment introduced by operation shuffling in cryptographic counter measures. In applications such as fuzzing and IDS, the most prevalent method is based on pattern matching~\cite{liang2021practical, sperl2019side}. However, variations in background processes from one execution to another can lead to misalignment issues. This challenge is thoroughly demonstrated in~\cite{liang2021practical}, where the authors aim to detect intrusions. They propose a solution called dynamic window matching to address this problem, but their approach still relies on pattern matching, which involves a heavy computational effort to search through the entire power trace for specific patterns. Moreover, if profiling the same target is not feasible, finding the correct point to align the traces becomes impractical.

With the advancement of ML and Deep Neural Networks (DNNs), these techniques have proven to be highly effective tools for signal processing. Researchers apply them to various \psc applications, either as classifiers or as pure signal processing tools for different purposes. For example, researchers use Convolutional Neural Networks (CNN) as a classifier for breaking cryptographic algorithms~\cite{hettwer2020encoding, picek2023sok}. Additionally, they have been used for power trace alignment~\cite{picek2018performance, kim2019make}. Furthermore, autoencoders as an unsupervised ML model are used for noise reduction of power traces~\cite{wu2020remove, kwon2021non}. Although ML-based models are efficient, they are data-driven. Meaning their performance depends heavily on the quality and representativeness of their training datasets. However, in applications where profiling is not feasible or where a dataset representing semi-real-world power traces is unavailable, these models struggle to generalize effectively limiting their applicability in such scenarios.


Moreover, ML models assume that the training and evaluation datasets follow an Identical and Independent Distribution (IID). This means that, for the ML model to perform effectively during an attack, the attacker must access to the same firmware (or firmware with a similar distribution of instructions) during the training phase. If this condition is not met, the trained ML model may not perform well in the attack phase. The issue of differing distributions between the training and evaluation datasets is known as Covariate Shift in the ML domain, and it remains an unresolved problem. All the studies mentioned in Table~\ref{tab:disassembly_compare} use datasets with the same distribution during both the profiling and attack phases, thereby avoiding the covariate shift problem. However, these attacks are not applicable in scenarios where the attacker cannot access to the firmware. As a result, none of the existing \psc-based disassemblers can effectively operate in black-box scenarios, rendering them inefficient in real-world cases where such access is unavailable.


\paragraph{\textbf{Experiment}}
To illustrate the challenge of covariate shift, we developed a simple random assembly code generator, similar to the approach proposed in \cite{narimani2207novel}, but this time tailored for the ARM Cortex-M4. Using this tool, we generate five random programs with uniformly distributed instructions. Additionally, we select the \texttt{AES128} encryption algorithm from tiny-AES-C \cite{tinyaes} as a real-world program. We conduct our experiments on the ideal STM32F3 target, executing all six programs (the five random ones and the AES128) while collecting power traces. In a manner similar to the experiment described in Section~\ref{sec:challenges:sysec:probing}, we train a MLP classifier with two hidden layers on four of the five random programs. We then evaluate the classifier twice: once on the fifth random program and once on the AES128 program. The accuracy drops significantly from 81\% to 41\%, demonstrating the impact of covariate shift. The drastic drop in accuracy during the second experiment occurres because the distribution of instructions in the AES encryption algorithm is not uniform, highlighting the covariate shift problem. In the Appendix~\ref{sec:appendix:CM2}, Figure~\ref{fig:CovariateShift_CM} displays the confusion matrix for these two scenarios. 

\subsection{Run-time Overhead}
\label{sec:challenges:syssec:realtime}

Repeating measurements for a single input is a common method to reduce noise in \psc attacks. However, in real-time-constrained applications such as fuzzing or intrusion detection, the practicability of the approach is limited due to the time-consuming nature of multiple executions and the requirement to reset the target to the same state each time. Therefore, execution per second is a limiting factor. In the domain of \psc-guided fuzzing, we only find one notable study, \cite{sperl2019side}, which theoretically implements fuzzing in a white-box scenario. We consider this study theoretical because it relies on previously recorded power traces instead of recording power traces and guiding the fuzzer in real-time. Additionally, even if power profiling is feasible in a white-box scenario, it is not possible in black-box systems as we do not have the firmware as the ground truth. Moreover, the paper does not provide details on execution time or a comparison with a blind fuzzer. This omission is significant because if a \psc-assisted fuzzer is slower than a blind fuzzer, its use may not be time-efficient. As a result, the challenge of effectively using \pscs in real-time applications such as fuzzing remains unaddressed, highlighting a gap in current research.


\subsection{Application Specific Challenge}
\label{sec:challenges:syssec:appspecific}

Apart from the mentioned challenges, there are a few application-specific challenges in the \psc. Specifically, we discuss crash detection, crash clustering, and cross-architecture generalization in fuzzing applications. 


\subsubsection{Crash Detection and Crash Clustering in Fuzzing:}
In fuzzing, crash detection is crucial. A reboot is a common crash signal this typically involves a fault handler that manages unexpected behavior. Faults vary, including CPU and software faults. Different embedded devices may run bare-metal-firmware or an RTOS, leading to varied implementations of fault handlers. This variability makes profiling even more challenging. Identifying the specific fault handler involved can provide valuable insights into the type of crash and its root cause. In these scenarios, particularly in black-box environments, there is often no debugging interface available, making crash detection and root cause analysis a rigorous task.


\subsubsection{Cross Architecture CFG Reconstruction:}
Another significant challenge for CFG analysis or IDS applications is their effectiveness across different architectures. Profiling-based approaches face a major hurdle in generalizing across architectures, as various architectures exhibit distinct leakage patterns due to differences in design, such as the number of pipeline stages. Most existing research has evaluated their methods on a single target or within a specific family of targets (as seen in Table~\ref{tab:disassembly_compare}). Consequently, the challenge of evaluating these methods across different architectures remains an open issue for CFG and IDS-related applications.


\begin{figure}[!t]
    \centering
    \includegraphics[width=7cm]{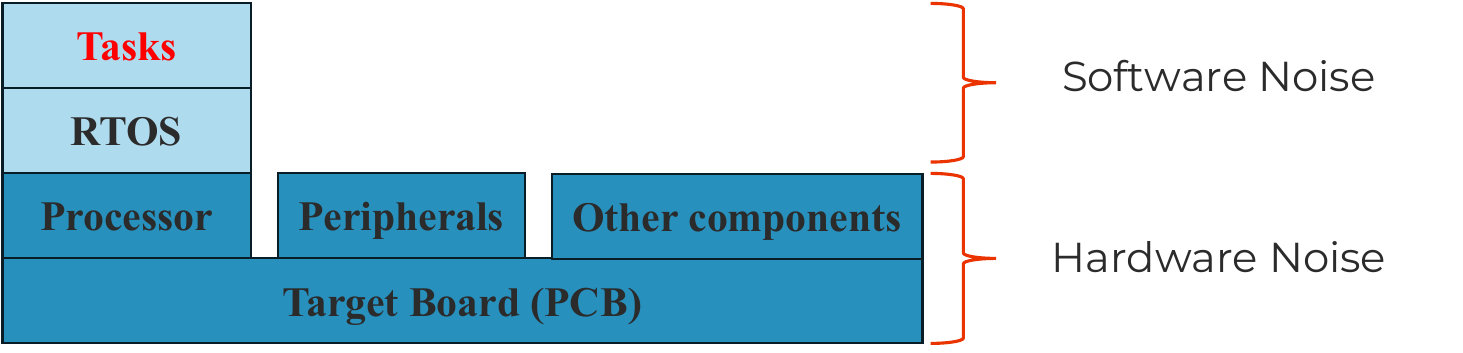}
    \caption{Software Noise vs Hardware Noise
    }
    \label{fig:SwN_vs_HwN}
\end{figure}
\subsection{Hardware noise vs. Software noise}

In this section, we discussed the challenges of using \psc in system security. From a high-level overview, we group the source of these challenges in two subgroups (Figure~\ref{fig:SwN_vs_HwN}).

\begin{itemize}
    \item
    \textbf{Hardware Noise:}
    This includes the noise generated by the PCB, peripherals, and components on it.
    \item
    \textbf{Software Noise:}
    This includes system-level interferences such as other processes, interrupts, and system calls. Software noise is the major challenge in using \psc in system security applications.
\end{itemize}

\section{Solutions}
\label{sec:solution}

This section discusses our proposed solution to address the raised challenges. First, we explain our method in detail and then design experiments to show its efficiency. Besides, we discuss the addressed challenge case by case, as mentioned in Section~\ref{sec:challenges:sysec}.

\subsection{Control Flow Fingerprinting Using \psc}
\label{sec:solution:CFG:fingerprint}

\psc was previously used for code execution tracking~\cite{liang2021practical}, but the mentioned challenges affect the measurements and reduce the accuracy in different applications; most are unsuitable for cross-device scenarios. In this section, we propose our algorithm based on wavelet transform to distinguish different code executions from power traces. 
As discussed in Section~\ref{sec:challenges:syssec:signalprocessing}, using ML models for signal processing is not desired here as they suffer from the covariate shift problem and are unsuitable for black-box applications.

Before explaining the details of the algorithm, we discuss what we will fingerprint. The typical power consumption of a code execution depends on the instructions and the data. Among instructions, branch instructions are the more interesting ones as if the branch is taken; not only should the current code execution stop, but the pipeline should also be cleared, and the program counter (PC) should change. These actions leave a distinguishable pattern in power trace as the amount of consumed power changes drastically. Consequently, these drastic changes lead to high-frequency changes in power traces. Therefore, power consumption leaks information about the CFG and can be used to track the code execution. However, \psc signals are noise-sensitive, and typically, multiple measurements (varies from ten to thousands)~\cite{hajra2013snr} are required to reduce the noise. This raises challenges in using \pscs in real-time applications.

\subsubsection{Frequency Domain Fingerprinting}
\label{sec:solution:CFG:fingerprint:alg}
\begin{table}[!t]
\captionsetup{font=footnotesize}
\footnotesize
    \centering
    \begin{tabular}{|c|c|c|c|}
         \hline
         Target & Accuracy & False positive & Drastic PC changes \\
         \hline
         AES 128 & 100\% & 10.82\% & 862 \\
         \hline
         SHA 256 & 100\% & 11.30\% & 841 \\
         \hline
         LZ4 & 100\% & 6.27\% & 2266 \\
         \hline
    \end{tabular}
    \caption{The accuracy and false positive rate of the wavelet-based control flow fingerprinting are reported in this table. Also, the number of detected branches against the number of existing branches are shown in last column.}
    \label{tab:wavelet_fingerprint}
\end{table}
With the identified source of information leakage for CFG fingerprinting, we now introduce our signal-processing algorithm.
Assuming that $x(t)$ is a measured raw power trace, first, a wavelet transform is applied to the power trace to extract the frequency domain representation, and the result is a $n*m$ matrix:
\begin{equation*}
W_x(t,a) =
\begin{bmatrix}
    g_{00} & g_{01} & . & g_{0m} \\
    g_{10} & g_{11} & . & g_{1m} \\
    . & . & ... & . \\
    g_{n0} & g_{n1} & . & g_{nm}
\end{bmatrix}
\end{equation*}
Where $a$ is the frequency scale of the wavelet. In this matrix, each row represents the wavelet transform for a fixed $a$ (daughter wavelet), and each column shows different scales in a single time slot. In this matrix, the values greater than a threshold $\Theta$ show the high-frequency changes with corresponding time slots. We filter the remaining values as follows:
\begin{equation*}
    \Omega^{\prime\prime}\ =\ \operatorname*{argmax}_t\ |W_x(t,a)|\ >\ \Theta
\end{equation*}
Where $\Omega^{\prime\prime}$ is a list of possible locations for high-frequency changes. Also, $\Theta$ is defined as follows:
\begin{equation*}
    \Theta\ =\ \theta\ *\ \max{|W_x(t,a)|}
\end{equation*}
Where $\theta$ is a threshold parameter that is set by the user. $\Omega^{\prime\prime}$ may have duplicate values for the same time slot because these high-frequency changes are observable in multiple frequencies as the wavelet transform filters the signal in a range of frequencies. We call these duplicate values
\textbf{spectral duplicates} and they are usually repeated across the same time slot but in different rows (usually neighboring frequencies). Therefore, we eliminate these duplicate values:
\begin{equation*}
    \Omega^\prime\ =\ Unique(\Omega^{\prime\prime})
\end{equation*}
\begin{figure}[t]
    \centering
    \includegraphics[width=5.8cm]{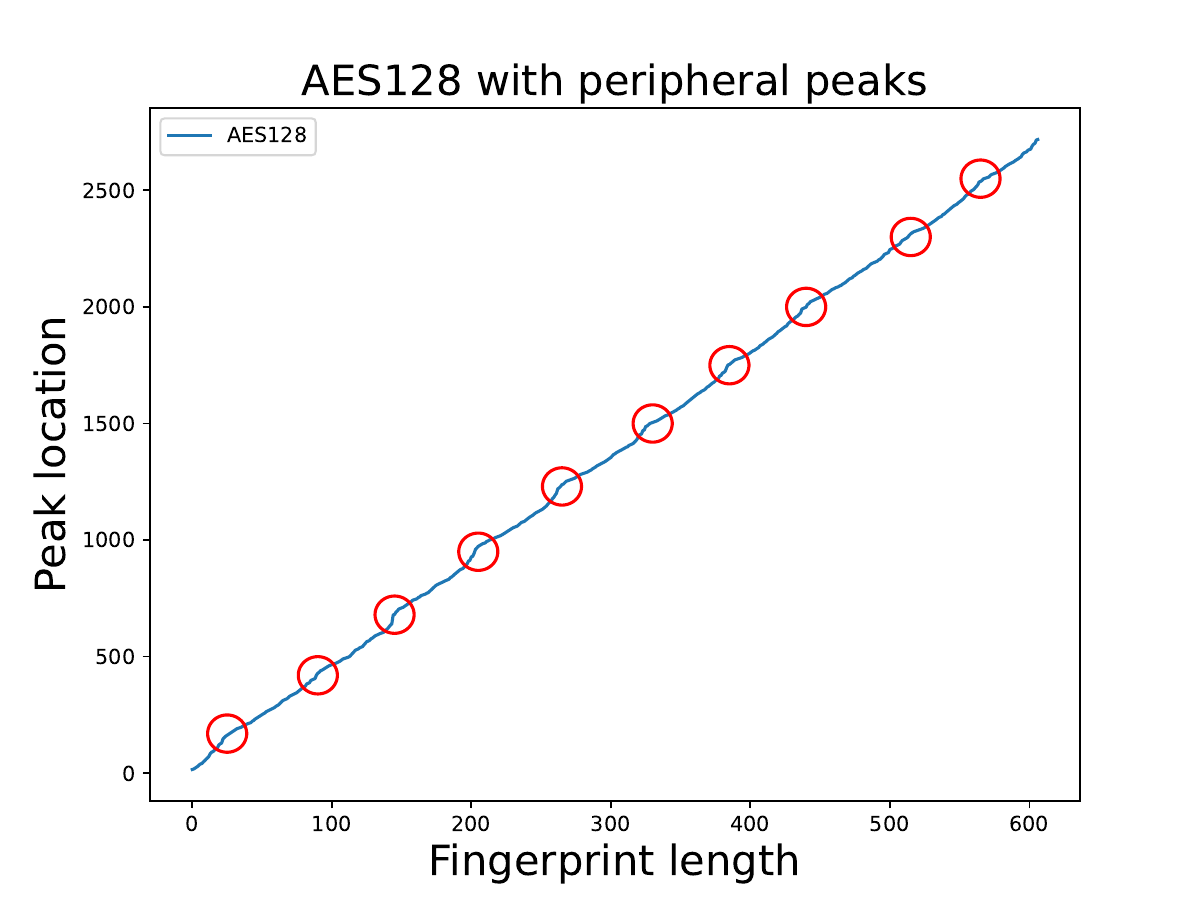}
    \caption{The fingerprint for AES128 in the presence of peripheral peaks, with ten rounds highlighted.}
    \label{fig:Wavelet_aes_pp}
\end{figure}
Where $\mathrm{\Omega^\prime}$ is a set of candidates representing time location (time slot) for high-frequency changes. To turn $\mathrm{\Omega^\prime}$ to a vector that fingerprints the program flow, we need to transform these values to instruction locations. Hence, we can use two approaches: 1) clock reconstruction or 2) division by sample per clock. Clock reconstruction provides a higher level of accuracy. However, assuming the attacker has access to the clock signal is a strong assumption. If the processor uses PLL to modify the external crystal's clock frequency, calculating the PLL multiplier values becomes impossible in a black-box system. In this case, we take the second approach and assume that the attacker can estimate the number of samples per clock, which is a reasonable assumption when using a power trace.
As a result, the values in $\mathrm{\Omega^\prime}$ are divided by the number of samples per clock. Here, another step is applied to eliminate the duplicate values that may exist due to having different samples from the same clock cycle in $\mathrm{\Omega^\prime}$. This mostly happens in cases where the sampling frequency is high enough that the size of the daughter wavelet $\psi_a(t)$ is similar to or smaller than a clock cycle. We call these duplicate values \textbf{temporal duplicates}. The final set of values that are represented as a vector $\mathrm{\Omega}$ is a unique identifier of the high-frequency changes in the signal $x(t)$.
\begin{equation*}
    \Omega\ =\ Unique(\{\omega\ |\ \Omega^\prime :\ \omega*\frac{F_s}{clk}\})
\end{equation*}
The most important advantage and a major contribution of our algorithm is that the proposed method does not need any information about the target firmware implementation, as we do not need to perform profiling on the same device. To demonstrate the practicality of our algorithm, we will show that an attacker can use it to extract information from a single execution run (without profiling the same device), making it a proper choice for real-time applications.
\paragraph{\textbf{Experiment}}
To evaluate the efficiency of our wavelet-based control flow fingerprinting algorithm with a single execution, we must first test our concept on a real-world microcontroller. Also, a thorough assessment of the accuracy and false positives is required. Therefore, we choose three different programs, namely \texttt{tinyaes128}~\cite{tinyaes}, \texttt{tinysha256}~\cite{sha256}, \texttt{lz4}~\cite{lz4}, and assess our algorithm's efficiency on them. For all three target programs, we set the threshold $\mathrm{\theta}$ to 0.3, the sampling frequency was 3.125 GS/s, and the target processor runs on an 8 MHz clock. We ran these programs on an STM32F3 board with a soldered shunt resistor.
Table~\ref{tab:wavelet_fingerprint} illustrates the result of the experiment. Accuracy is the number of detected severe PC changes compared to the existing ones, and false positive is the rate of falsely detected changes compared to the existing ones. As the results indicate, the proposed method can identify changes in the PC and represent it as a unique fingerprint for the code execution.

Our proposed signal-processing algorithm performs well in fingerprinting the control flow of a black-box target. The most crucial point about our algorithm is that it can work with a single power trace. However, the downside of this method is the high resolution (3.125 GS/s) and high amount of computations. Also, as mentioned in Section~\ref{sec:challenges:syssec:noise}, peripheral access can cause massive voltage peaks that may affect the accuracy of our method. Later, we evaluate our method in such cases and try to improve it.

\subsection{Probing}
\label{sec:solution:probing}
\begin{figure}[!t]
    \centering
    \includegraphics[width=6cm]{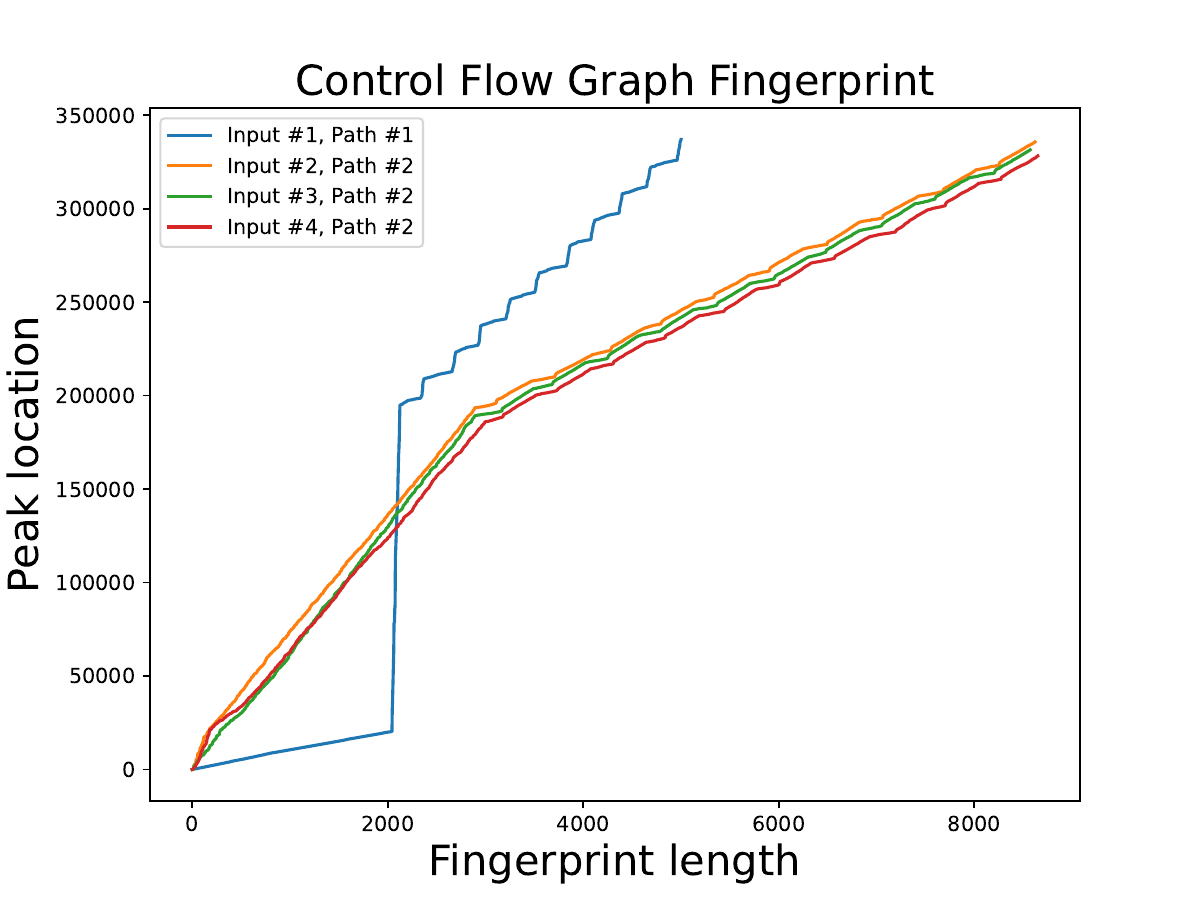}
    \caption{CFG fingerprinting for an ini parser. All inputs have the same execution time with different fingerprints.}
    \label{fig:CFG_ini}
\end{figure}
A good probing point should be as close as possible to the main processor to avoid having extra noise from other components. However, this is impossible in an off-the-shelf device with many components and peripherals. This challenge is addressed by eliminating peripherals and having a single processor in the CW board. This solution seems efficient in applications where the firmware binary image without peripheral interaction is available. So, the attacker can load it on the CW and apply the \psc attack. However, firmware is usually unavailable for most embedded targets, especially in black-box scenarios.
To solve this issue, we propose using a custom-designed interposer with considerations for \psc measurements. An interposer is an electrical device acting as an interface between two connections, allowing a third connection to monitor the main signal. For high-speed signals, the interposer handles the impedance matching to avoid signal disturbance. Here, the designed interposer should cut the power lines and put a shunt resistor in place for proper power sampling.
Then, the main processor should be desoldered to put the interposer on the main PCB and then solder the target processor on the interposer. Although this may require a single interposer for each package type, this can solve the \psc measurement problem for real-world embedded devices.

\subsection{Noise}
\label{sec:solution:noise}
\begin{figure}[!t]
    \centering
    \includegraphics[width=6cm]{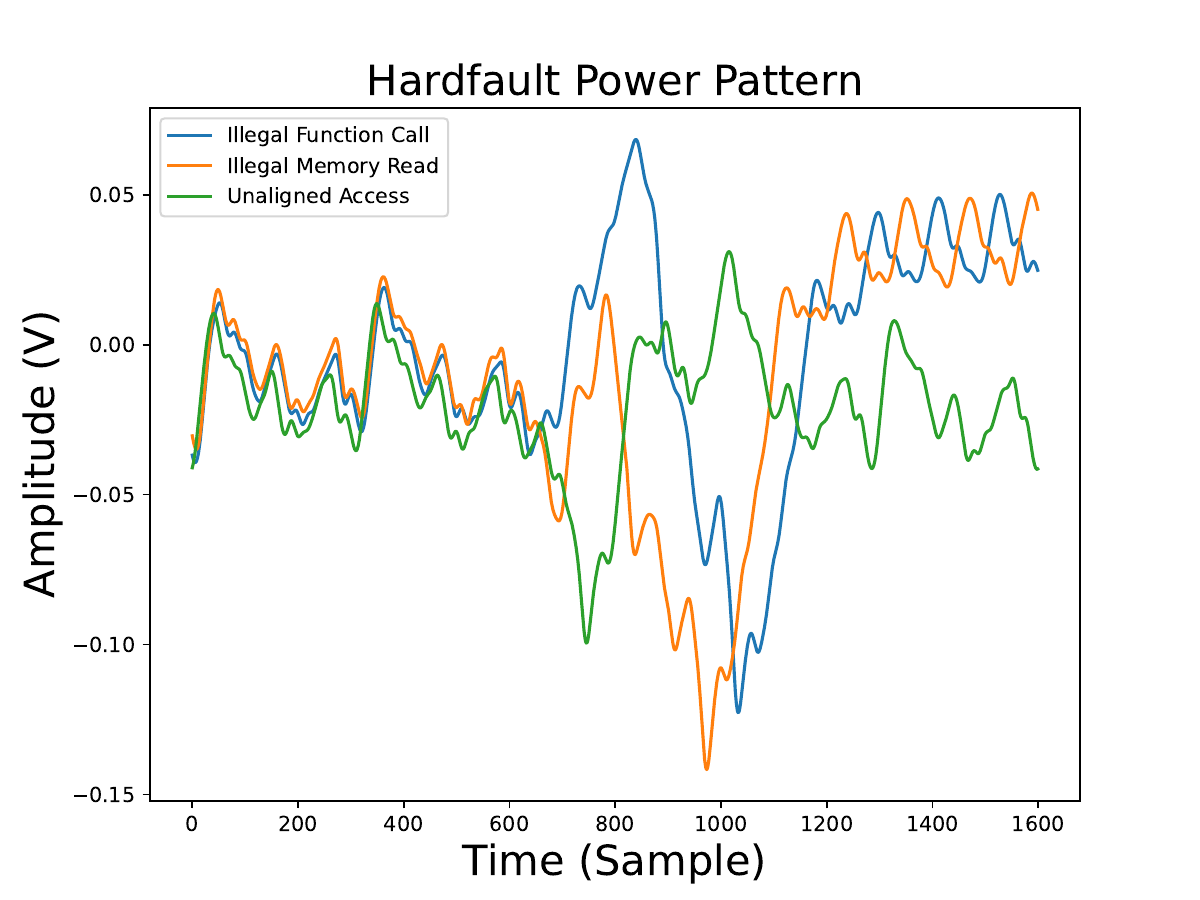}
    \caption{Power pattern for hard-faults of a STM32F3 and their comparison (Jitter is removed by a low-pass filter).
    }
    \label{fig:Crash_Cluster}
\end{figure}
%

    
This subsection explains the effect of peripheral interaction on the wavelet-based fingerprinting algorithm. As the peripheral interaction leaves huge peaks on the power traces, this can significantly affect the $\theta$ value in our algorithm, making it sensitive to noise. We need to analyze the behavior of the firmware functionality during these peaks to find the solution. Basically, the processor is either sending data, receiving data, or both during these peaks, meaning there is no interesting information about the internal firmware in these cycles. Therefore, removing the part of the power trace corresponding to these peaks causes no damage to the resulting fingerprint. Hence, our algorithm detects these large peaks and removes them. To show this solution actually works, some peripheral interactions are intentionally injected into the AES128 programs from subsection \ref{sec:solution:CFG:fingerprint}. Figure~\ref{fig:Wavelet_aes_pp} shows the fingerprint for the AES128 with peripheral peaks in the power trace. As it is observable, removing these peaks from the power trace makes our approach work. This proves the robustness of our method against peripheral peaks.


\subsection{Signal Processing}
\label{sec:solution:signal:processing}

This section evaluates our proposed frequency-domain fingerprinting technique from a signal-processing perspective. Traditional pattern-matching methods are not robust to time-shift, which are common in embedded firmware running on an RTOS, and addressing this issue requires additional signal processing to locate the correct position in the trace. In contrast, our method does not require matching the entire power trace, as the relevant information is already contained within the fingerprint. Figure~\ref{fig:CFG_ini} shows the fingerprints for an \texttt{ini file parser}~\cite{ini}. While all inputs result in the same execution time, the fingerprints differ, indicating different execution paths (potentially new basic blocks). A similarity metric such as Mean Square Error (MSE) or the correlation coefficient can easily detect these differences. The power traces were collected from an STM32F3 microcontroller, running with an 8 MHz clock and no peripheral interaction, at a sampling rate of 50 MS/s, allowing real-time processing.

\subsection{Run-time Overhead}
\label{sec:solution:realtime}
Existing \psc-based methods are unsuitable for real-time use due to their reliance on multiple traces to improve the SNR. In contrast, wavelet-based fingerprinting can extract accurate fingerprints from a single measurement without noise reduction techniques. Although false positives are inevitable in black-box scenarios, they remain below 12\% (As shown in Table~\ref{tab:wavelet_fingerprint}), making our method viable for real-time applications like IDS and fuzzing.

\subsection{Crash Detection and Crash Clustering}
\label{sec:solution:application:specific:crash}
\begin{figure}[!t]
    \centering
    \subfloat[\centering AVR Fingerprints]{{\includegraphics[width=4.4cm]{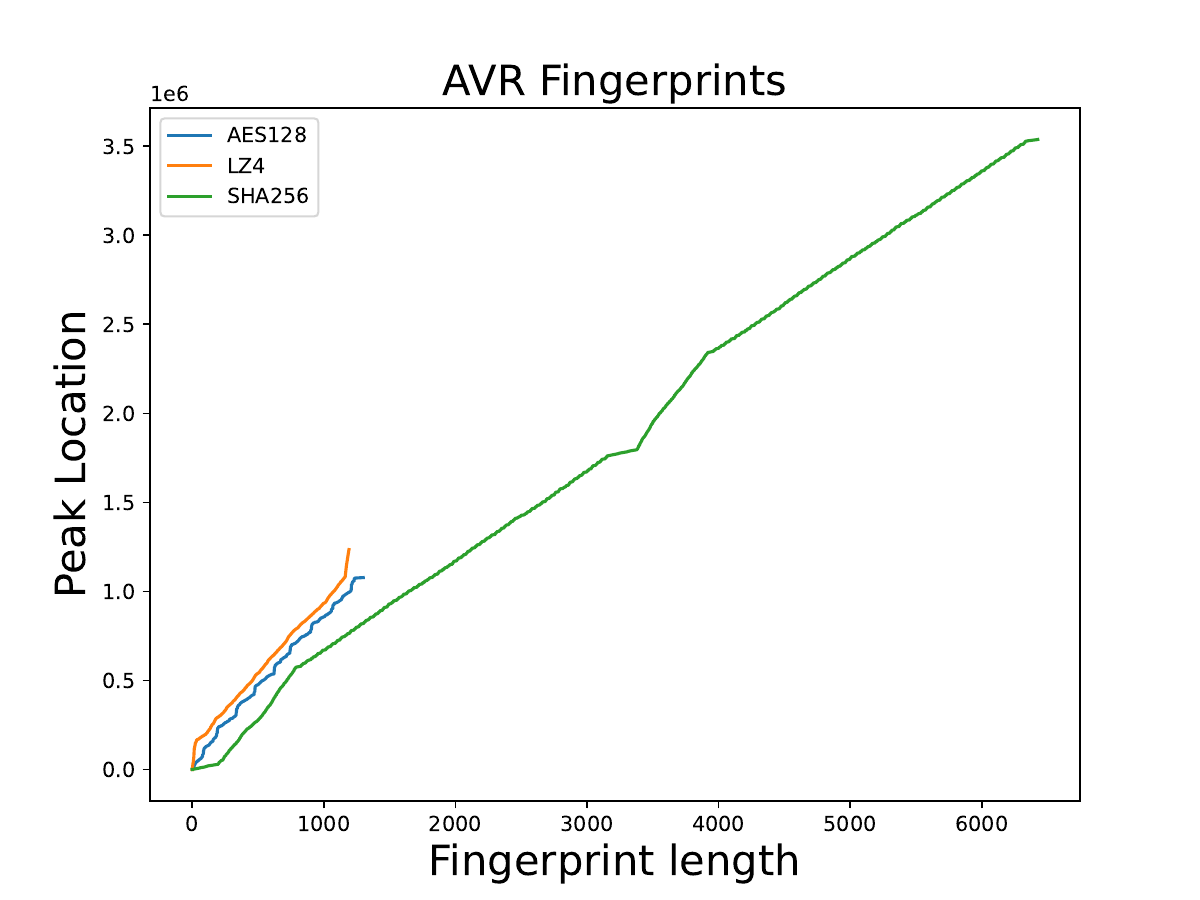}}}
    \subfloat[\centering ARM Fingerprints]{{\includegraphics[width=4.4cm]{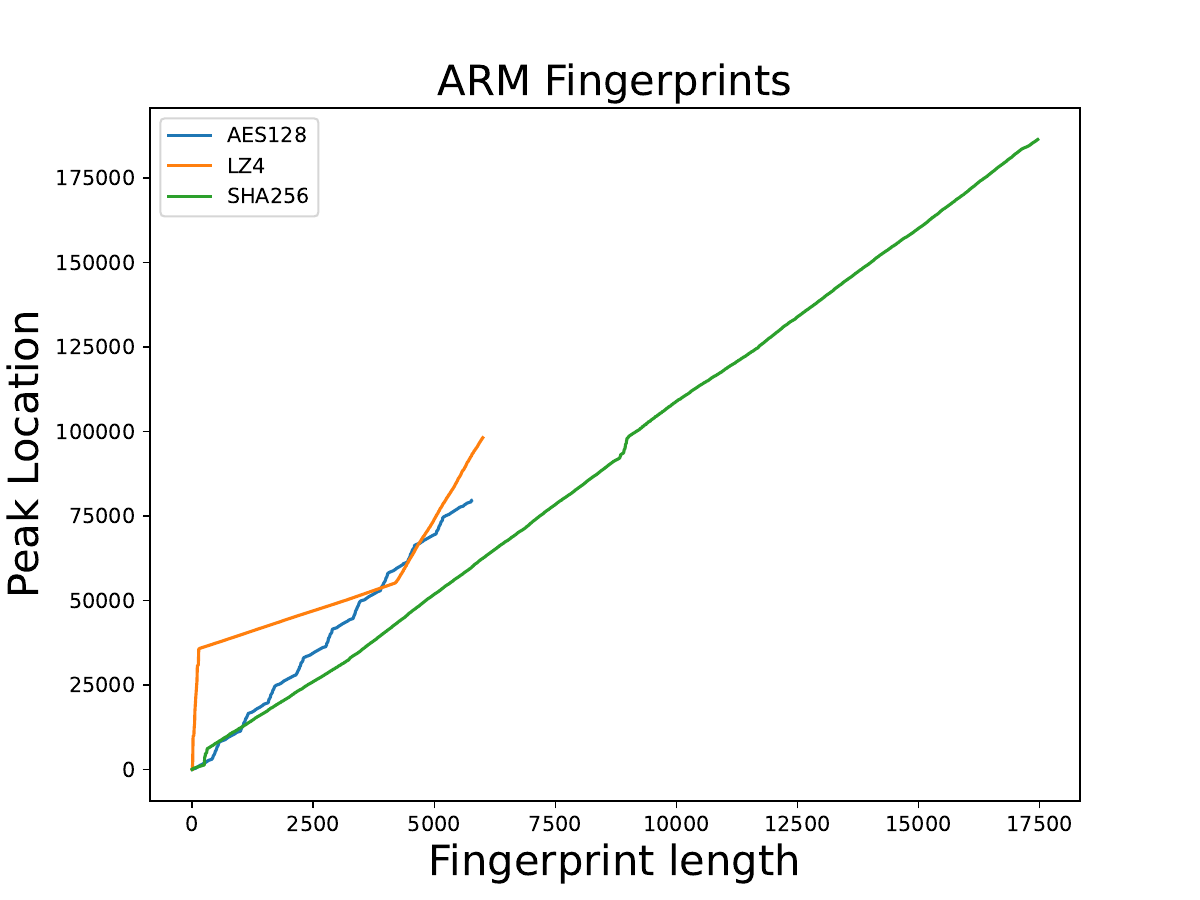}}}
    \caption{Cross Architecture Fingerprints}
    \label{fig:Cross_Arc}
\end{figure}
To show that different types of faults may have different power trace patterns, we trigger actual crashes in a processor and monitor the power consumption signals. To do so, we chose the STM32F (Cortex-M4) board with a shunt resistor soldered on it. The clock frequency for the microcontroller is 8 MHz, and the sampling frequency of the oscilloscope is 625 MS/s.

We trigger the faults in the middle of a function, executing a mock workload to simulate the fault occurring during computation. This leads the processor to execute a handler for the fault. Fingerprinting the fault handler itself might yield data that could be used to detect or cluster crashes, as different types of failures might cause differences in control flow in the fault handler. However, we observe that the firmware of embedded devices often does not contain error-handling logic in the fault handlers. This is probably because one of these faults occurring is a sign that something went wrong in an unexpected way necessitating a reset of the device. The default fault handler in projects setup using the STM32CubeIDE~\cite{stm32CubeIDE} is just a spinning loop. For the use case of crash detection, detecting this spinning loop is sufficient. The behavior and, thus, power trace of the hardware before executing the software fault handler differs depending on the fault type.
Figure~\ref{fig:Crash_Cluster} shows the power pattern for three hard-faults namely \texttt{Illegal Function Call} (function pointer to an invalid address), \texttt{Illegal Memory Read}, and \texttt{Unaligned Access}. The power patterns are distinct for each fault. Depending on the fault, a large peak may be a rising or falling, while the pattern preceding the fault remains the same across all three cases. Note that these power patterns correspond to the fault's hardware handling before triggering the software handler.

When applied to software testing, this detectable behavioral difference can be used to cluster crashes by the type of fault they cause. For example, if an input triggers a crash due to a function call at an invalid address, and modifying part of that input results in a crash caused by an invalid instruction, we can infer that the crashing input likely corrupts a function pointer. This clustering helps identify the root causes of crashes, enabling debugging and vulnerability identification.

\subsection{Cross Architecture}
\label{sec:solution:application:specific:cross}

As explained previously, cross-architecture generalization is a bottleneck for \psc applications. To show that our method can overcome this challenge, we implement the three programs from subsection \ref{sec:solution:CFG:fingerprint} on a different architecture. We choose the \texttt{ATXMEGA128}, an \texttt{AVR} microcontroller with a soldered shunt resistor on the PCB. To run our experiment, we set the sampling frequency of the oscilloscope to 62.5 MS/s and probe the power on the soldered shunt resistor on the PCB. Figure~\ref{fig:Cross_Arc} shows the extracted fingerprints and the different execution paths for different programs, while the fingerprints are similar to the ones that we executed on the \texttt{ARM} architecture. This proves the cross-architecture generalizability of our method and the uniqueness of the extracted fingerprint regardless of the underlying hardware.

\section{Evaluation}
\label{sec:evaluation}

This section implements and tests our solutions on real-world targets. Furthermore, we discuss the extra signal processing steps required to handle our target noise.



\subsection{Real-World Target Fingerprinting}
\label{sec:evaluation:real:world:fingerprint}

\begin{figure}[!t]
    \centering
    \subfloat[\centering Raw Ublox power trace]{{\includegraphics[width=4.2cm]{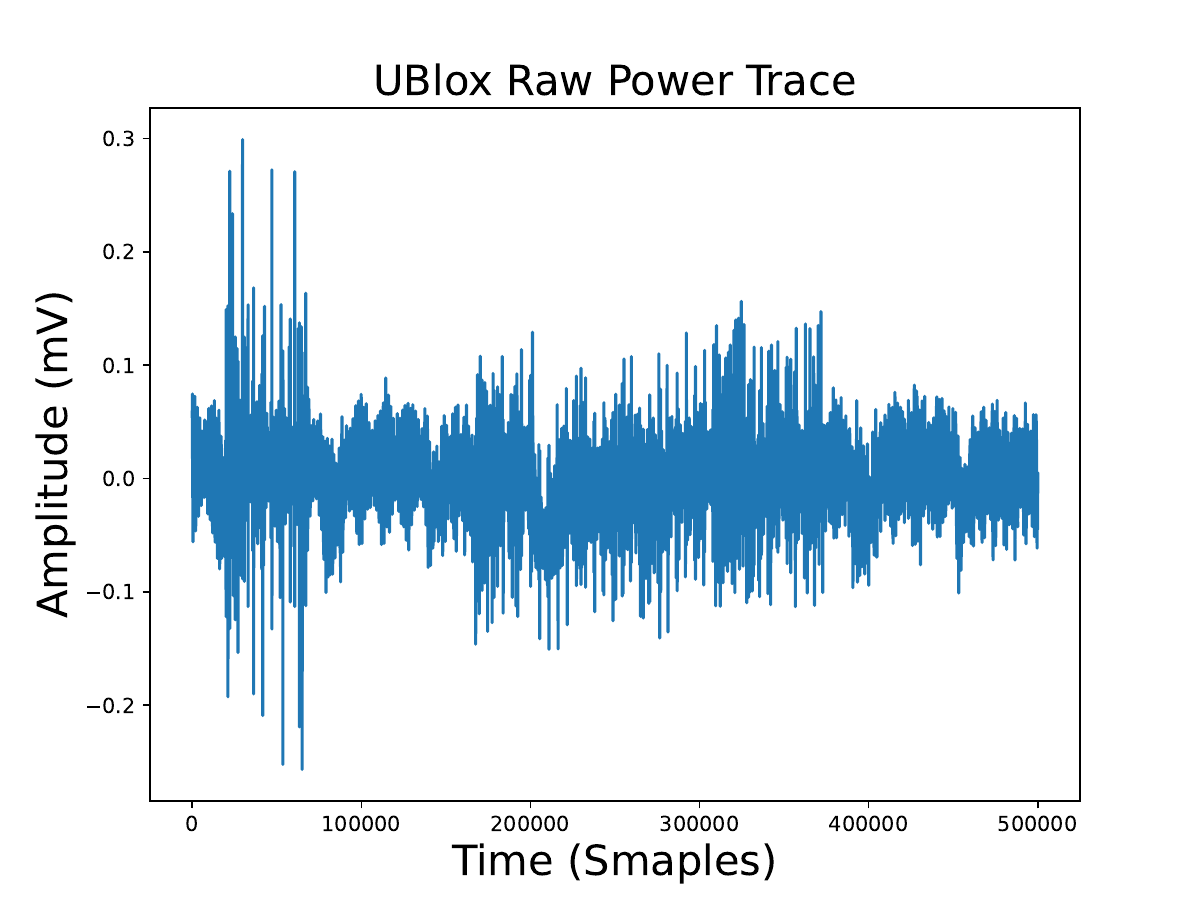}}}
    \subfloat[\centering UBlox power trace after applying a high-pass filter]{{\includegraphics[width=4.2cm]{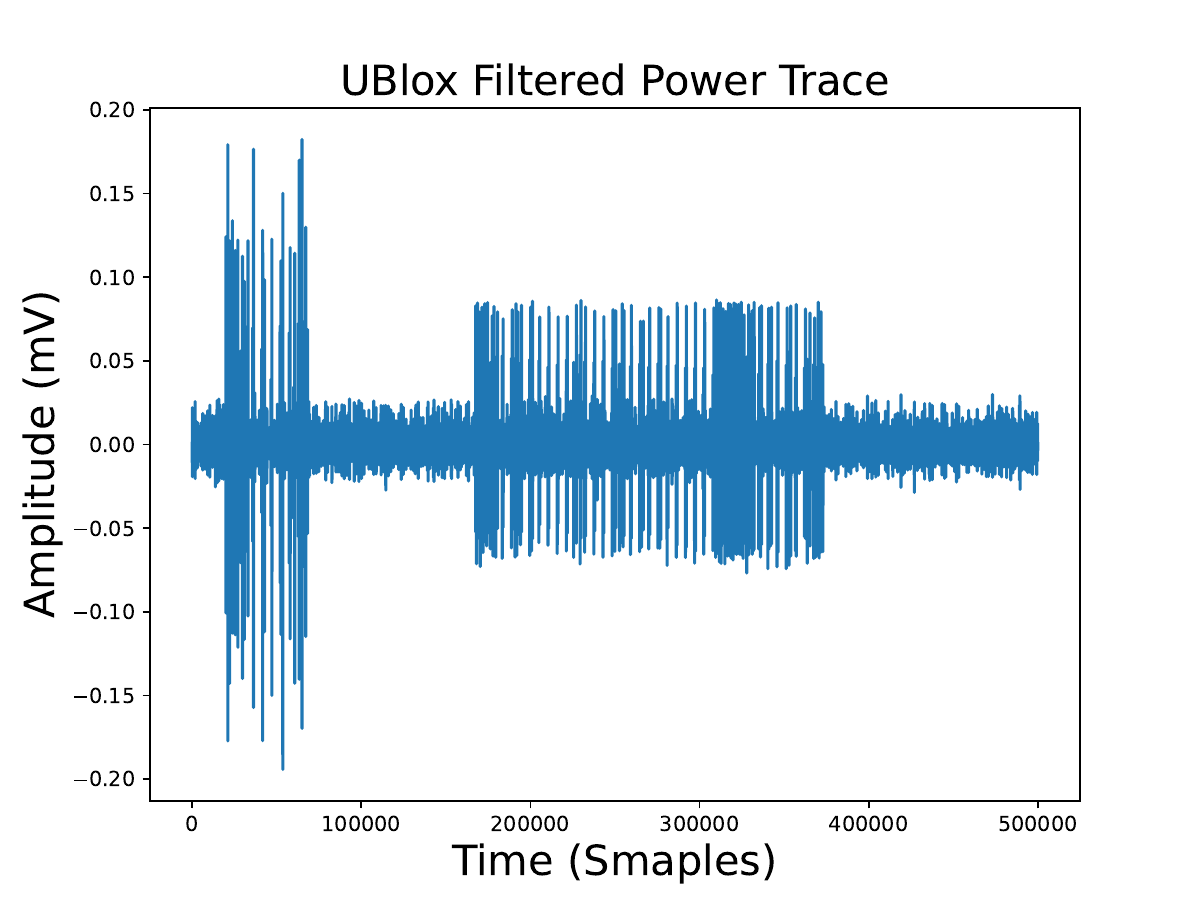}}}
    \caption{UBlox power traces before and after filtering}
    \label{fig:UBlox_power_traces}
\end{figure}
To evaluate the efficiency of the proposed wavelet-based fingerprinting method on a real-world target, we conducted a gray-box test using the UBlox GNSS module without prior knowledge of its firmware or hardware.
As shown in Figure~\ref{fig:UBlox_Unboxed} in the Appendix~\ref{sec:appendix:UBlox}, the module has a RF noise-stop shield. We removed the shield to access the main processor pins for a closer probing point. The figure illustrates two identical microprocessors without shunt resistors. We then probe the power pins of the second processor, as highlighted in the figure, since the RF signals are routed to it after passing through a mixer. We pass the \texttt{UBX-CFG-PRT} with different payloads.

Figure~\ref{fig:UBlox_power_traces} (a) shows one of the power traces for these commands. As illustrated, the peripheral peaks are detectable, and low-frequency noise is also observable, complicating fingerprinting. The other components on the board cause this noise. Fingerprinting this signal will not produce robust results as the noise highly affects the wavelet transform. However, recalling the wavelet-based fingerprinting subsection, we know that the PC-related information is encoded in high frequencies. Thus, a high-pass filter removes the low-frequency noise without compromising the desired information. We use a Butterworth high-pass filter of degree five, and the resulting signal is shown in Figure~\ref{fig:UBlox_power_traces} (b).

\begin{figure}[!t]
    \centering
    \includegraphics[width=5.8cm]{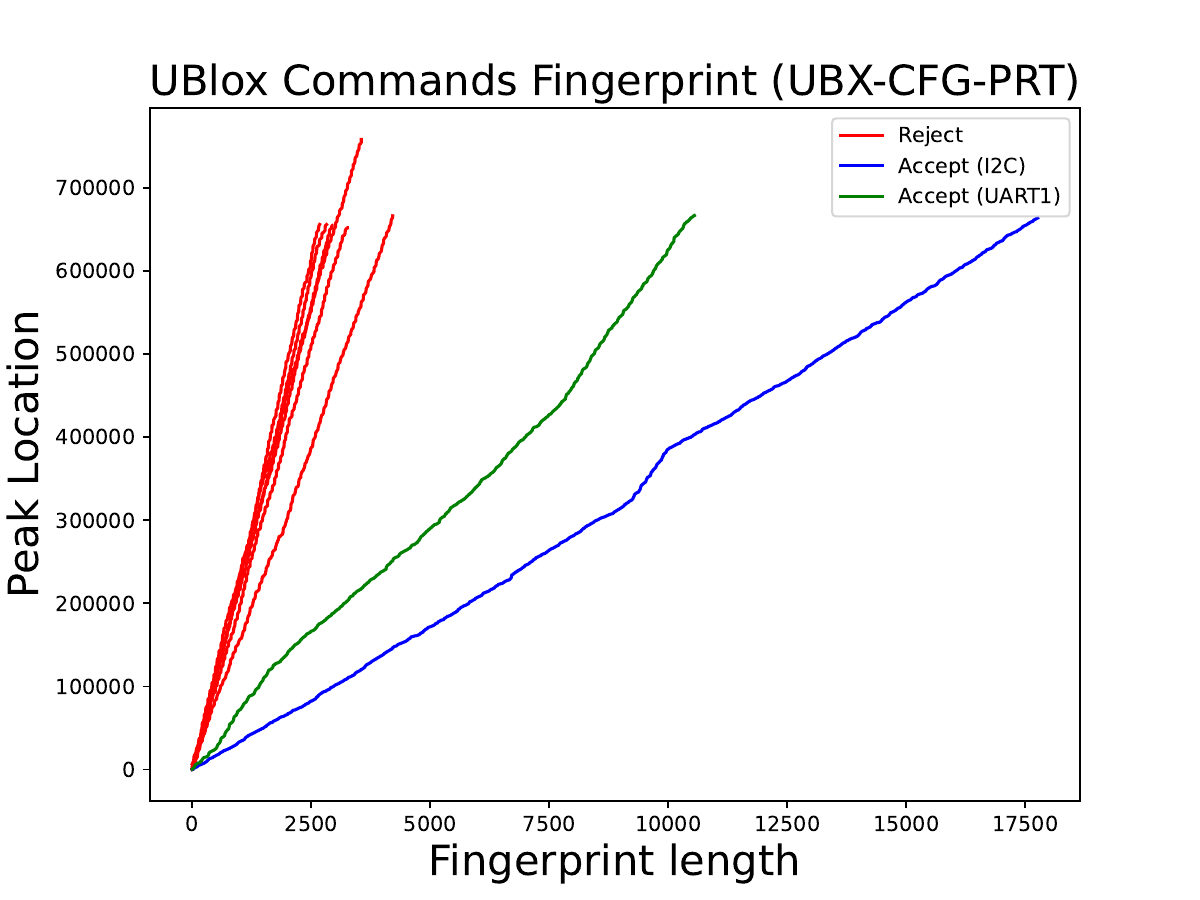}
    \caption{The extracted fingerprints for different commands from a UBlox module}
    \label{fig:UBlox_command_fingerprints}
\end{figure}
With the low-frequency noise removed, fingerprinting the power traces becomes feasible. As previously mentioned, peripheral peaks can disturb the fingerprinting algorithm. However, they do not contain information relevant to our analysis, as they occur when the processor sends or receives data. Instead, our focus is on the processes applied to the received packets. Therefore, the power trace is clipped, and only the section between the sending and receiving peripheral peaks is used for fingerprinting. Figure~\ref{fig:UBlox_command_fingerprints} shows the fingerprints, where the differences in the execution path of rejected and accepted payloads are distinguishable.

\subsection{Peripheral Interaction}
\label{sec:evaluation:peripheral}

\begin{figure}
    \centering
    \subfloat[\centering Raw Data]{{\includegraphics[width=4.4cm]{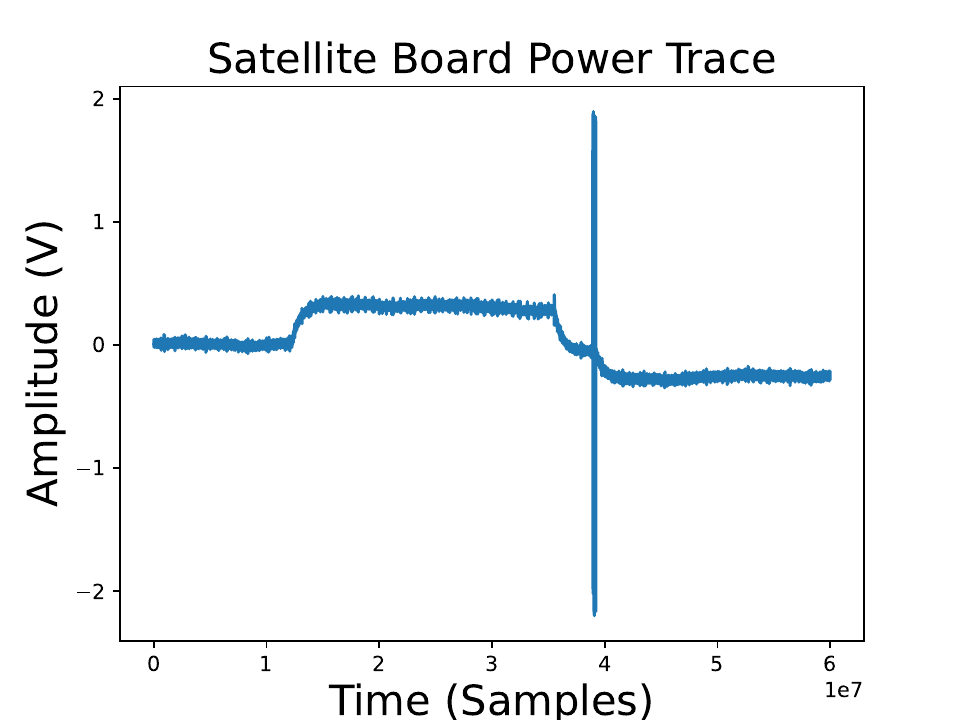}}}
    \subfloat[\centering Group of peaks]{{\includegraphics[width=4.4cm]{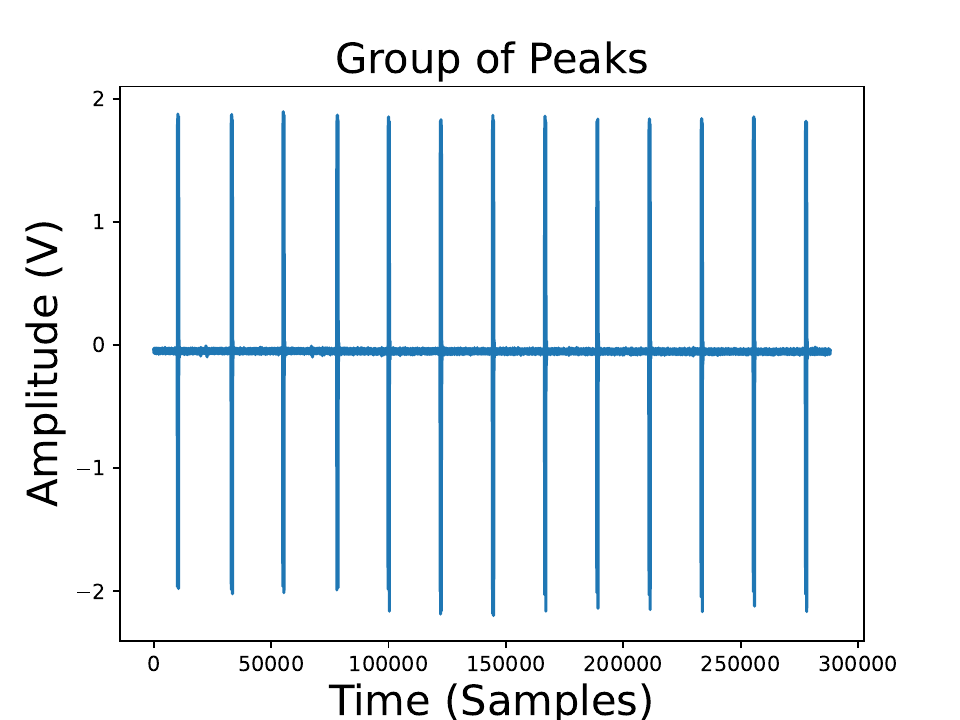}}}

    \subfloat[\centering One byte data transfer]{{\includegraphics[width=4.4cm]{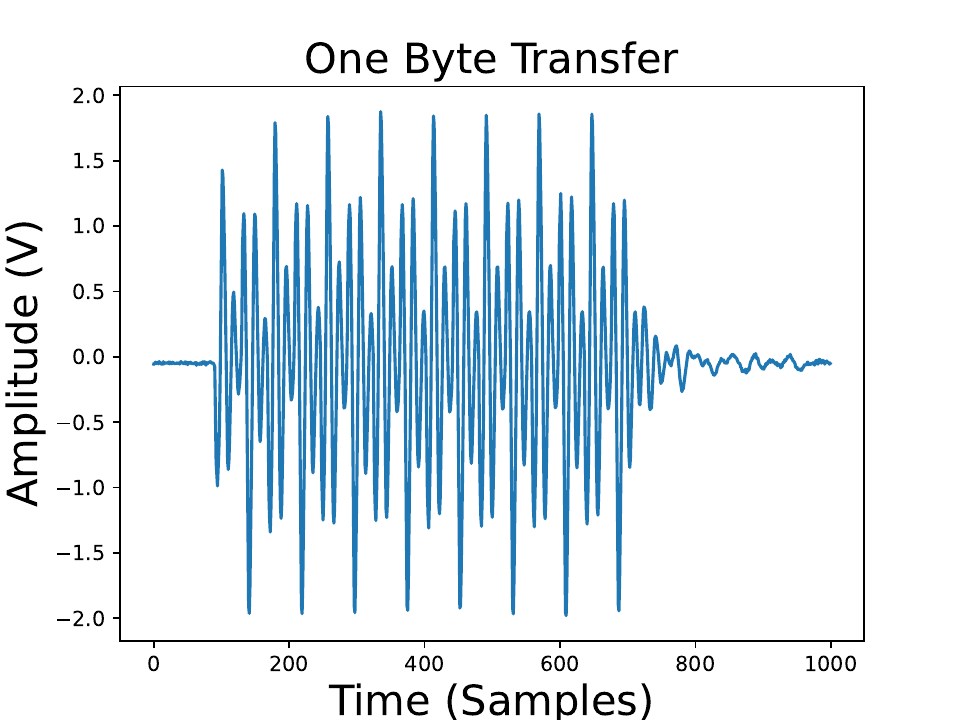}}}
    \subfloat[\centering Decoded data]{{\includegraphics[width=4.4cm]{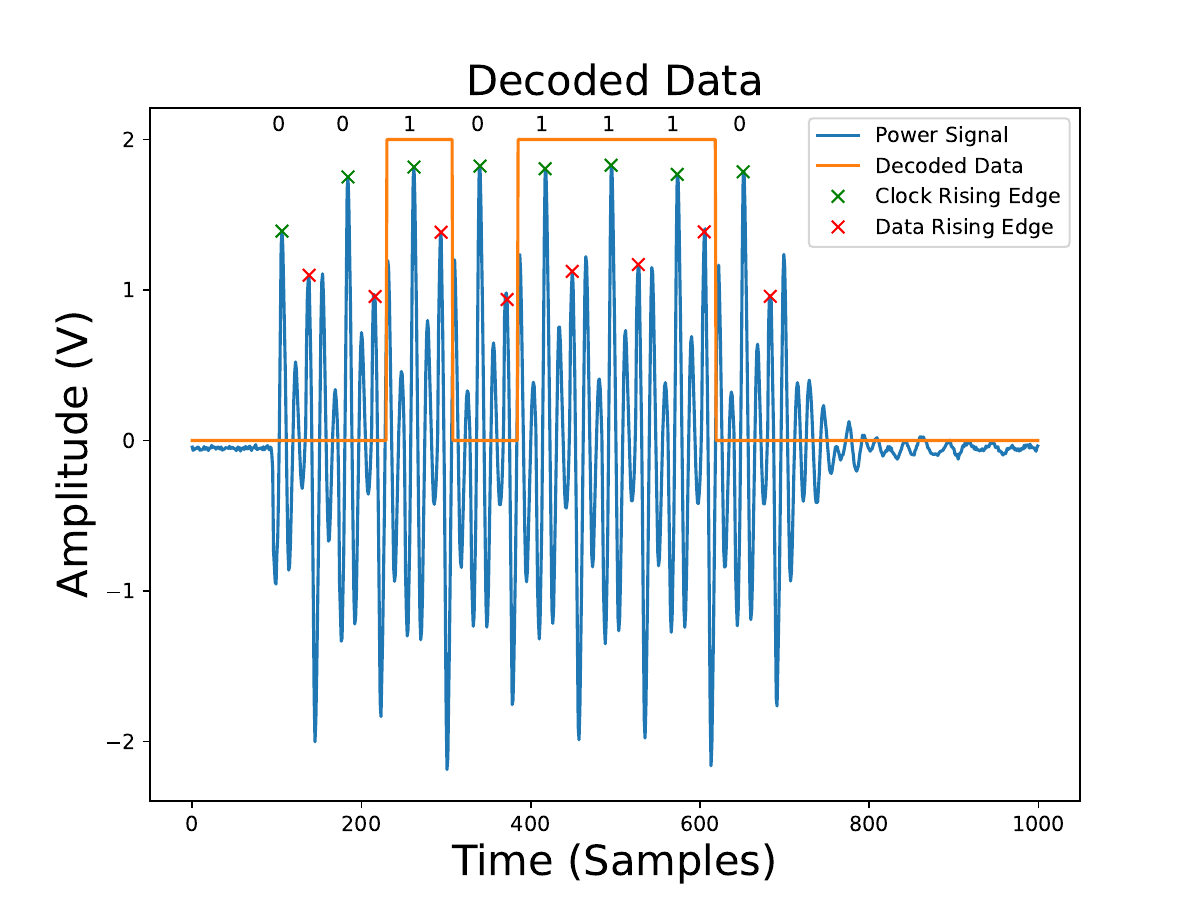}}}
    
    \caption{Decoding the data transfer of a satellite board communicating through SPI}
    \label{fig:Peripheral_satellite}
\end{figure}
\begin{figure*}[!t]
    \centering
    \subfloat[\centering Raw Power Trace]{{\includegraphics[width=4.4cm]{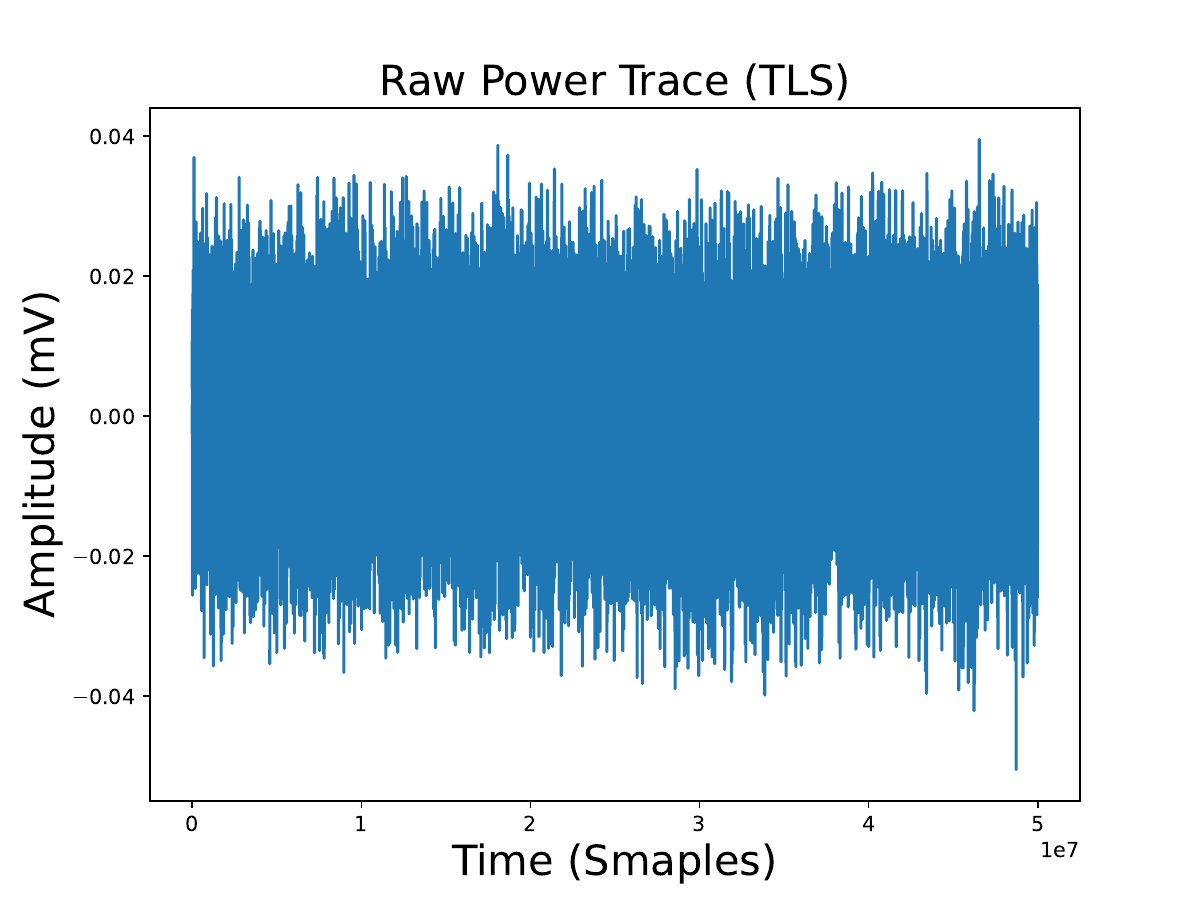}}}
    \subfloat[\centering Filtered Power Trace]{{\includegraphics[width=4.4cm]{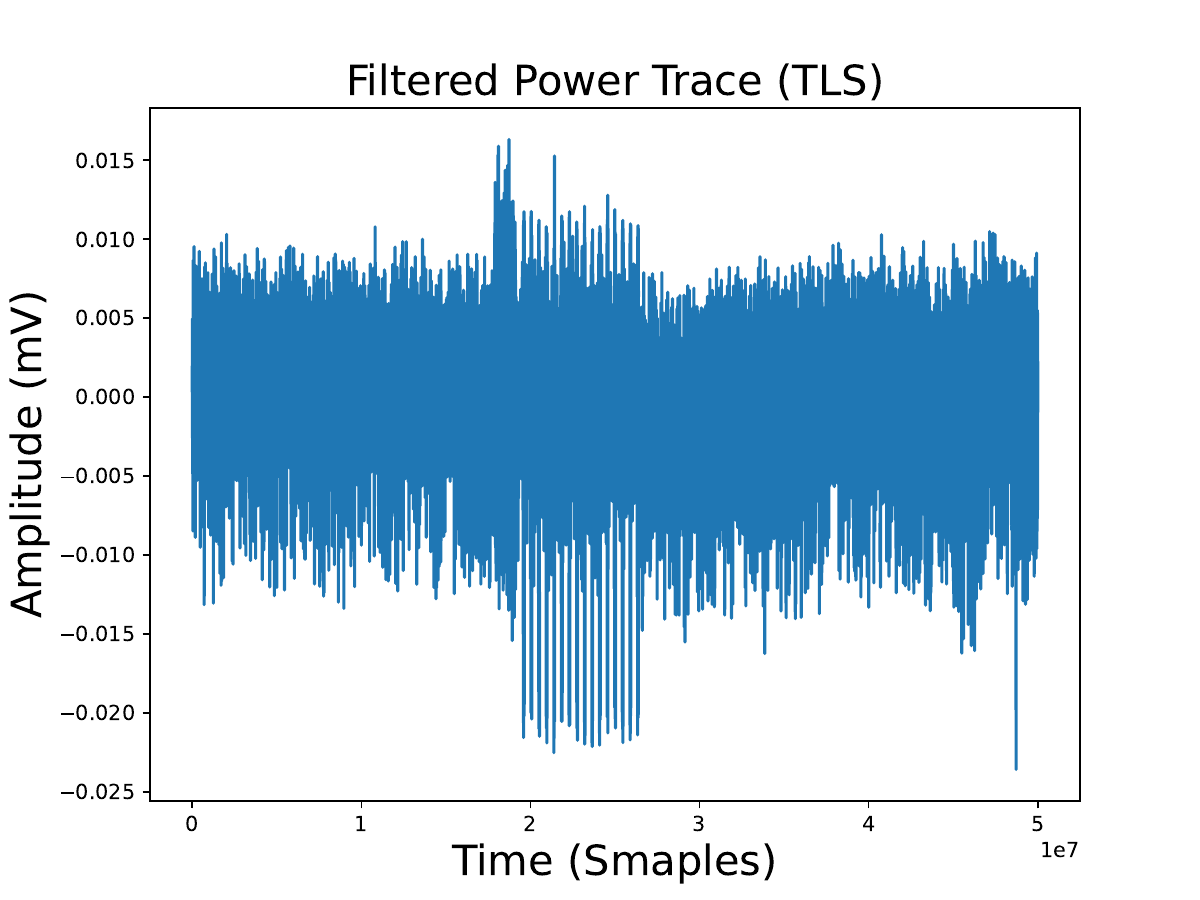}}}
    \subfloat[\centering Fingerprint]{{\includegraphics[width=4.4cm]{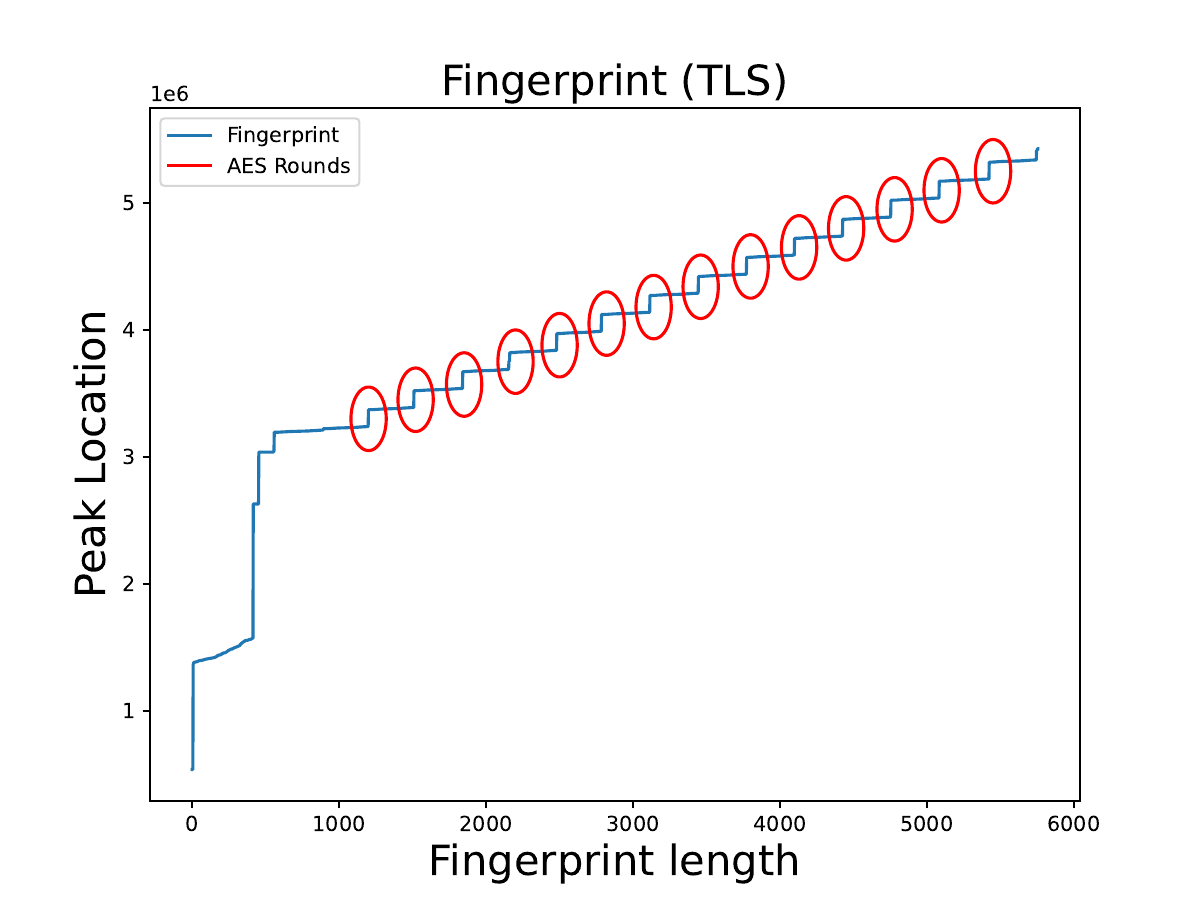}}}
    
    \caption{Power trace and extracted fingerprint for the secure MQTT sensor/actuator sample from Zephyr project.
    }
    \label{fig:Sofware_Composition_Analysis}
\end{figure*}

In this subsection, we evaluate the peripheral interaction detection on a real-world target. High peaks in power consumption are observable during signal transitions on an external interface, where the significant current is either consumed or released. According to Ohm's law ($\bm{R}=\frac{\bm{V}}{\bm{I}}$, where R is the resistance, V is the voltage, and I is the current), during a rising edge, a high current is consumed briefly, leading to a voltage drop in the power trace (assuming constant resistance). During a falling edge, the current previously consumed is released, increasing the observed voltage. Using this algorithm, we detect both rising and falling edges on the data lines, enabling the decoding of data from the power trace.

We use the satellite board to evaluate our algorithm in a real-world scenario. The board includes a Ferroelectric RAM (FRAM), and our goal is to decode the data transferred between the processor and the FRAM. For this experiment, we set the sampling frequency to 1.25 Gs/s. Figure~\ref{fig:Peripheral_satellite} (a) shows the power trace captured by probing the board's power lines without hardware manipulation. As shown in the figure, the high peaks indicate peripheral interactions. Cutting the power trace in the interaction area, we observe a group of peaks (Figure~\ref{fig:Peripheral_satellite} (b)). Zooming on one of the peaks, we see a signal with eight large peaks and smaller peaks in between (Figure~\ref{fig:Peripheral_satellite} (c)). This pattern belongs to the Serial Peripheral Interface (SPI), as eight repeated high peaks represent the clock cycles for a byte transfer. Figure~\ref{fig:Peripheral_satellite} (d) shows the decoded data using the abovementioned algorithm. In SPI, data sampling by the receiver is synchronized with the rising edge of the clock (green peaks), meaning that data transitions occur on the falling edge of the clock (red peaks). By classifying the red peaks into three groups, we distinguish rising edges, falling edges, and those without transitions.

\subsection{Software Composition Analysis}
\label{sec:evaluation:composition}

In this section, we evaluate the proposed wavelet-based algorithm to detect a AES algorithm in real-world targets with real-world embedded RTOS running on them. We designed two experiments namely low-end and high-end.

\subsubsection{Low-End}
As a low-end target board, we choose the Nucleo board mentioned in Section~\ref{sec:challenges:sysec:expsetup}. To simulate a realistic application, we select the \emph{Secure MQTT Sensor/Actuator}~\cite{zephyrSecureMqttSample} sample from the Zephyr OS project~\cite{zephyr} and build it for the Nucleo board without modification.

Without customization for \psc measurement or any injected trigger signals, we record the power trace of the processor for two seconds. For our application, it should be more than enough to have a successful communication. The sampling frequency of the oscilloscope is set to 25 MS/s, and the probing point is a power pin on the main processor.

Figure~\ref{fig:Sofware_Composition_Analysis} (a) shows the Raw power trace recorded from the target board. The effect of noise is observable on the signal as no data is detectable from the raw power trace. Same as in the previous subsection, we apply a high-pass filter to reduce low-frequency noise. Figure~\ref{fig:Sofware_Composition_Analysis} (b) shows the denoised signal and the encryption parts that consume more power are distinguishable in the filtered trace. In the last step, we apply our fingerprinting algorithm on the filtered signal and plot the resulting fingerprint in Figure~\ref{fig:Sofware_Composition_Analysis} (c). The rounds of AES encryption are visible and plotted inside red circles in the figure. As the encryption is AES 256 here, it has more rounds than the AES 128 used in Section~\ref{sec:solution}.

\subsubsection{High-End}
\begin{figure}[!t]
    \centering
    \includegraphics[width=5.8cm]{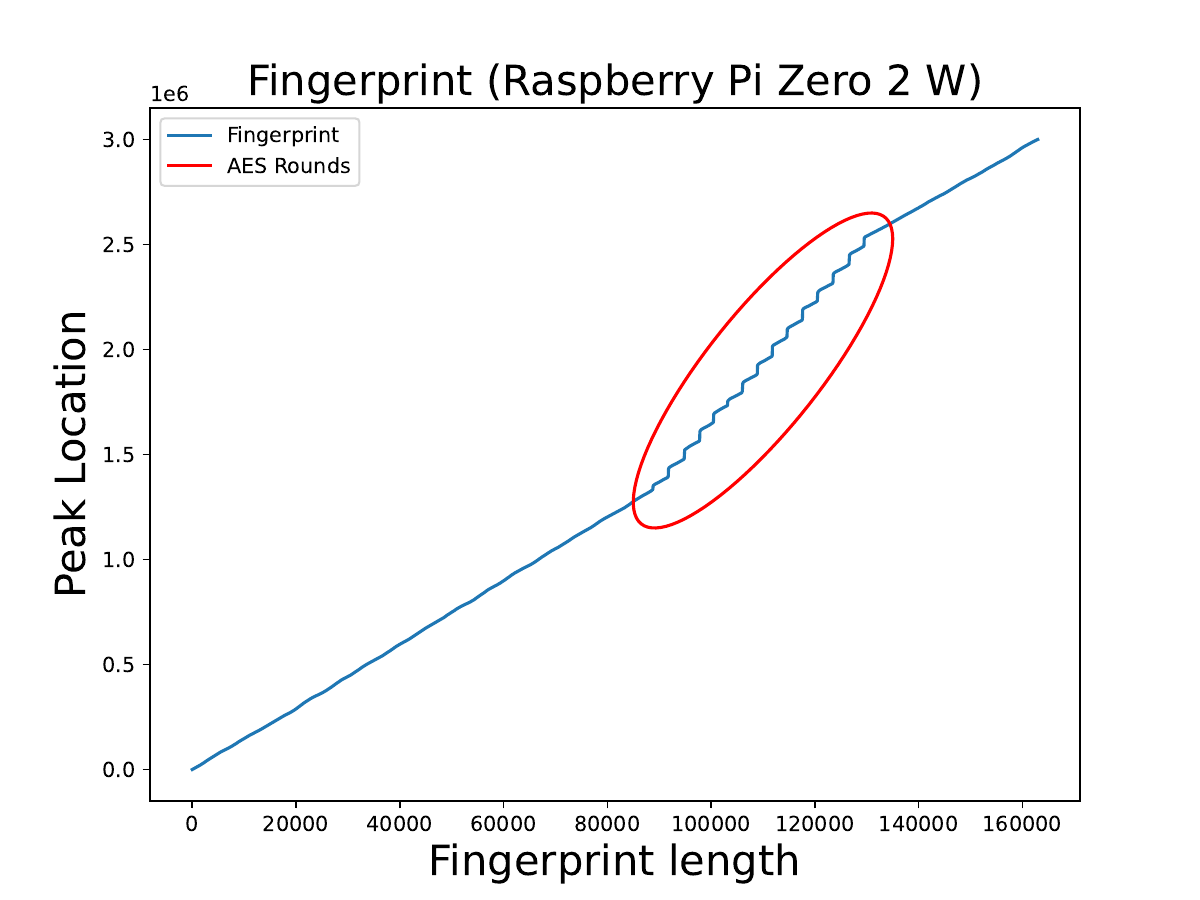}
    \caption{Fingerprint of an AES running on the Raspberry Pi Zero 2 W.}
    \label{fig:RasPi}
\end{figure}

As a high-end target, we choose the Raspberry Pi Zero running the Raspbian kernel on it. The sampling frequency of the oscilloscope is set to 250 MS/s. We establish a TLS connection on it and measure the power trace during execution. After applying a low-pass filter, we apply the fingerprinting algorithm. Figure~\ref{fig:RasPi} shows the extracted fingerprint, and the AES rounds are again visible in this plot.

Detecting the AES algorithm in a real-world target shows that the proposed fingerprinting algorithm is able to perform across different development boards with different architectures.

\section{Discussion}
\label{sec:discussion}

In this paper, we proposed a fingerprinting method to eliminate dependencies on specific hardware or firmware. Although signal processing algorithms may introduce performance bottlenecks, this can be mitigated or resolved with more optimized hardware. In our prototype, we used an oscilloscope for sampling and transferred the power traces to our host machine in real-time via a direct Ethernet connection. The wavelet transform was optimized to run in parallel on an Nvidia 3070 Mobile GPU. The time required for data transfer and processing is dependent on the size of the power trace. However, using an oscilloscope connected to a host machine in a real-time scenario can limit processing speed, particularly with large power traces. We aim to address this limitation in our future work by implementing our algorithm on an FPGA that could be a more viable solution. FPGA implementation would eliminate the need to transfer large power traces to an external host, thereby enabling real-time processing and significantly enhancing the speed and efficiency of our algorithm.

\section{Conclusion}
\label{sec:conclusion}
In this paper, we systematically examined the application of \psc attacks in the embedded system security context. While \psc attacks have successfully extracted cryptographic keys, expanding their use to broader embedded system applications poses unique challenges. Our analysis identified key factors that reduce the effectiveness of traditional \psc techniques in real-world settings. To address these barriers in PSC applications, we introduced a novel signal processing method, particularly by eliminating the need for profiling the same target and a shunt resistor. Our experiments on real-world embedded devices demonstrate that this method is both effective and scalable, even with limited knowledge of the target device and firmware, without the need for any hardware modification. Our findings suggest that \psc attacks can be more widely applied in embedded system security by overcoming the discussed challenges. Future work can focus on refining these methods for broader applications and setups, ensuring that \psc becomes a critical tool for embedded systems security.

\bibliographystyle{IEEEtran}
\bibliography{ref}

\begin{thebibliography}{10}
\providecommand{\url}[1]{#1}
\csname url@samestyle\endcsname
\providecommand{\newblock}{\relax}
\providecommand{\bibinfo}[2]{#2}
\providecommand{\BIBentrySTDinterwordspacing}{\spaceskip=0pt\relax}
\providecommand{\BIBentryALTinterwordstretchfactor}{4}
\providecommand{\BIBentryALTinterwordspacing}{\spaceskip=\fontdimen2\font plus
\BIBentryALTinterwordstretchfactor\fontdimen3\font minus \fontdimen4\font\relax}
\providecommand{\BIBforeignlanguage}[2]{{%
\expandafter\ifx\csname l@#1\endcsname\relax
\typeout{** WARNING: IEEEtran.bst: No hyphenation pattern has been}%
\typeout{** loaded for the language `#1'. Using the pattern for}%
\typeout{** the default language instead.}%
\else
\language=\csname l@#1\endcsname
\fi
#2}}
\providecommand{\BIBdecl}{\relax}
\BIBdecl

\bibitem{chari2003template}
S.~Chari, J.~R. Rao, and P.~Rohatgi, ``Template attacks,'' in \emph{Cryptographic Hardware and Embedded Systems-CHES 2002: 4th International Workshop Redwood Shores, CA, USA, August 13--15, 2002 Revised Papers 4}.\hskip 1em plus 0.5em minus 0.4em\relax Springer, 2003, pp. 13--28.

\bibitem{brier2004correlation}
E.~Brier, C.~Clavier, and F.~Olivier, ``Correlation power analysis with a leakage model,'' in \emph{Cryptographic Hardware and Embedded Systems-CHES 2004: 6th International Workshop Cambridge, MA, USA, August 11-13, 2004. Proceedings 6}.\hskip 1em plus 0.5em minus 0.4em\relax Springer, 2004, pp. 16--29.

\bibitem{schindler2005stochastic}
W.~Schindler, K.~Lemke, and C.~Paar, ``A stochastic model for differential side channel cryptanalysis,'' in \emph{Cryptographic Hardware and Embedded Systems--CHES 2005: 7th International Workshop, Edinburgh, UK, August 29--September 1, 2005. Proceedings 7}.\hskip 1em plus 0.5em minus 0.4em\relax Springer, 2005, pp. 30--46.

\bibitem{gierlichs2008mutual}
B.~Gierlichs, L.~Batina, P.~Tuyls, and B.~Preneel, ``Mutual information analysis: A generic side-channel distinguisher,'' in \emph{International Workshop on Cryptographic Hardware and Embedded Systems}.\hskip 1em plus 0.5em minus 0.4em\relax Springer, 2008, pp. 426--442.

\bibitem{kocher1999differential}
P.~Kocher, J.~Jaffe, and B.~Jun, ``Differential power analysis,'' in \emph{Advances in Cryptology—CRYPTO’99: 19th Annual International Cryptology Conference Santa Barbara, California, USA, August 15--19, 1999 Proceedings 19}.\hskip 1em plus 0.5em minus 0.4em\relax Springer, 1999, pp. 388--397.

\bibitem{eisenbarth2010building}
T.~Eisenbarth, C.~Paar, and B.~Weghenkel, ``Building a side channel based disassembler,'' \emph{Transactions on computational science X: special issue on security in computing, part I}, pp. 78--99, 2010.

\bibitem{park2018power}
J.~Park, X.~Xu, Y.~Jin, D.~Forte, and M.~Tehranipoor, ``Power-based side-channel instruction-level disassembler,'' in \emph{Proceedings of the 55th Annual Design Automation Conference}, 2018, pp. 1--6.

\bibitem{narimani2021side}
P.~Narimani, M.~A. Akhaee, and S.~A. Habibi, ``Side-channel based disassembler for avr micro-controllers using convolutional neural networks,'' in \emph{2021 18th International ISC Conference on Information Security and Cryptology (ISCISC)}.\hskip 1em plus 0.5em minus 0.4em\relax IEEE, 2021, pp. 75--80.

\bibitem{liu2016code}
Y.~Liu, L.~Wei, Z.~Zhou, K.~Zhang, W.~Xu, and Q.~Xu, ``On code execution tracking via power side-channel,'' in \emph{ACM Conference on Computer and Communications Security (CCS)}, 2016, pp. 1019--1031.

\bibitem{sperl2019side}
P.~Sperl and K.~B{\"o}ttinger, ``Side-channel aware fuzzing,'' in \emph{Computer Security--ESORICS 2019: 24th European Symposium on Research in Computer Security, Luxembourg, September 23--27, 2019, Proceedings, Part I 24}.\hskip 1em plus 0.5em minus 0.4em\relax Springer, 2019, pp. 259--278.

\bibitem{han2022hiding}
Y.~Han, M.~Chan, Z.~Aref, N.~O. Tippenhauer, and S.~Zonouz, ``Hiding in plain sight? on the efficacy of power side $\{$Channel-Based$\}$ control flow monitoring,'' in \emph{31st USENIX Security Symposium (USENIX Security 22)}, 2022, pp. 661--678.

\bibitem{liang2021practical}
S.~Liang, X.~Peng, H.~J. Qi, S.~Zonouz, and R.~Beyah, ``A practical side-channel based intrusion detection system for additive manufacturing systems,'' in \emph{2021 IEEE 41st International Conference on Distributed Computing Systems (ICDCS)}.\hskip 1em plus 0.5em minus 0.4em\relax IEEE, 2021, pp. 1075--1087.

\bibitem{hori2012sasebo}
Y.~Hori, T.~Katashita, A.~Sasaki, and A.~Satoh, ``Sasebo-giii: A hardware security evaluation board equipped with a 28-nm fpga,'' in \emph{The 1st IEEE Global Conference on Consumer Electronics 2012}.\hskip 1em plus 0.5em minus 0.4em\relax IEEE, 2012, pp. 657--660.

\bibitem{dubey2020maskednet}
A.~Dubey, R.~Cammarota, and A.~Aysu, ``Maskednet: The first hardware inference engine aiming power side-channel protection,'' in \emph{2020 IEEE International Symposium on Hardware Oriented Security and Trust (HOST)}.\hskip 1em plus 0.5em minus 0.4em\relax IEEE, 2020, pp. 197--208.

\bibitem{msgna2014precise}
M.~Msgna, K.~Markantonakis, and K.~Mayes, ``Precise instruction-level side channel profiling of embedded processors,'' in \emph{Information Security Practice and Experience: 10th International Conference, ISPEC 2014, Fuzhou, China, May 5-8, 2014. Proceedings 10}.\hskip 1em plus 0.5em minus 0.4em\relax Springer, 2014, pp. 129--143.

\bibitem{jungmin2019leveraging}
P.~Jungmin, R.~Fahim, V.~Apostol, F.~Domenic, and T.~Mark, ``Leveraging side-channel information for disassembly and security,'' \emph{J. Emerg. Technol. Comput. Syst}, vol.~16, no.~1, 2019.

\bibitem{muench2018you}
M.~Muench, J.~Stijohann, F.~Kargl, A.~Francillon, and D.~Balzarotti, ``What you corrupt is not what you crash: Challenges in fuzzing embedded devices.'' in \emph{NDSS}, 2018.

\bibitem{paper-repo}
anonymous authors, ``Exploring power side-channel challenges in embedded systems security,'' \url{https://anonymous.4open.science/r/psc-sys-sec-0E54/}, 2024.

\bibitem{bracewell1989fourier}
R.~N. Bracewell, ``The fourier transform,'' \emph{Scientific American}, vol. 260, no.~6, pp. 86--95, 1989.

\bibitem{daubechies1990wavelet}
I.~Daubechies, ``The wavelet transform, time-frequency localization and signal analysis,'' \emph{IEEE transactions on information theory}, vol.~36, no.~5, pp. 961--1005, 1990.

\bibitem{ChipWhispererHusky}
{NewAE Technology Inc.}, ``{C}hip{W}hisperer-{H}usky,'' \url{https://rtfm.newae.com/Capture/ChipWhisperer-Husky/}, 2024.

\bibitem{hajra2014multivariate}
S.~Hajra and D.~Mukhopadhyay, ``Multivariate leakage model for improving non-profiling dpa on noisy power traces,'' in \emph{Information Security and Cryptology: 9th International Conference, Inscrypt 2013, Guangzhou, China, November 27-30, 2013, Revised Selected Papers 9}.\hskip 1em plus 0.5em minus 0.4em\relax Springer, 2014, pp. 325--342.

\bibitem{das2019x}
D.~Das, A.~Golder, J.~Danial, S.~Ghosh, A.~Raychowdhury, and S.~Sen, ``X-deepsca: Cross-device deep learning side channel attack,'' in \emph{Proceedings of the 56th Annual Design Automation Conference 2019}, 2019, pp. 1--6.

\bibitem{picek2017side}
S.~Picek, A.~Heuser, A.~Jovic, S.~A. Ludwig, S.~Guilley, D.~Jakobovic, and N.~Mentens, ``Side-channel analysis and machine learning: A practical perspective,'' in \emph{2017 International Joint Conference on Neural Networks (IJCNN)}.\hskip 1em plus 0.5em minus 0.4em\relax IEEE, 2017, pp. 4095--4102.

\bibitem{bruneau2015less}
N.~Bruneau, S.~Guilley, A.~Heuser, D.~Marion, and O.~Rioul, ``Less is more: dimensionality reduction from a theoretical perspective,'' in \emph{Cryptographic Hardware and Embedded Systems--CHES 2015: 17th International Workshop, Saint-Malo, France, September 13-16, 2015, Proceedings 17}.\hskip 1em plus 0.5em minus 0.4em\relax Springer, 2015, pp. 22--41.

\bibitem{guneysu2011generic}
T.~G{\"u}neysu and A.~Moradi, ``Generic side-channel countermeasures for reconfigurable devices,'' in \emph{International Workshop on Cryptographic Hardware and Embedded Systems}.\hskip 1em plus 0.5em minus 0.4em\relax Springer, 2011, pp. 33--48.

\bibitem{hettwer2020applications}
B.~Hettwer, S.~Gehrer, and T.~G{\"u}neysu, ``Applications of machine learning techniques in side-channel attacks: a survey,'' \emph{Journal of Cryptographic Engineering}, vol.~10, no.~2, pp. 135--162, 2020.

\bibitem{cagli2017convolutional}
E.~Cagli, C.~Dumas, and E.~Prouff, ``Convolutional neural networks with data augmentation against jitter-based countermeasures: Profiling attacks without pre-processing,'' in \emph{Cryptographic Hardware and Embedded Systems--CHES 2017: 19th International Conference, Taipei, Taiwan, September 25-28, 2017, Proceedings}.\hskip 1em plus 0.5em minus 0.4em\relax Springer, 2017, pp. 45--68.

\bibitem{van2022side}
J.~van Geest and I.~Buhan, ``A side-channel based disassembler for the arm-cortex m0,'' in \emph{International Conference on Applied Cryptography and Network Security}.\hskip 1em plus 0.5em minus 0.4em\relax Springer, 2022, pp. 183--199.

\bibitem{narimani2207novel}
P.~Narimani, S.~Habibi, and M.~A. Akhaee, ``A novel framework for dataset generation for profiling disassembly attacks using side-channel leakages and deep neural networks,'' \emph{arXiv preprint arXiv:2207.12068}, 2022.

\bibitem{cristiani2020bit}
V.~Cristiani, M.~Lecomte, and T.~Hiscock, ``A bit-level approach to side channel based disassembling,'' in \emph{Smart Card Research and Advanced Applications: 18th International Conference, CARDIS 2019, Prague, Czech Republic, November 11--13, 2019, Revised Selected Papers 18}.\hskip 1em plus 0.5em minus 0.4em\relax Springer, 2020, pp. 143--158.

\bibitem{han2017watch}
Y.~Han, S.~Etigowni, H.~Liu, S.~Zonouz, and A.~Petropulu, ``Watch me, but don't touch me! contactless control flow monitoring via electromagnetic emanations,'' in \emph{Proceedings of the 2017 ACM SIGSAC conference on computer and communications security}, 2017, pp. 1095--1108.

\bibitem{Nucleo144}
{STMicroelectronics}, ``{NUCLEO-144},'' \url{https://www.st.com/en/evaluation-tools/nucleo-f429zi.html}, 2017.

\bibitem{ubloxmodule}
{UBlox}, ``{UBlox ZED-F9P},'' \url{https://www.u-blox.com/en/product/zed-f9p-module}, 2024.

\bibitem{RasPi}
``{Raspberry Pi},'' \url{https://www.raspberrypi.com/products/raspberry-pi-zero-2-w/}, 2021.

\bibitem{katashita2012side}
T.~Katashita, Y.~Hori, H.~Sakane, and A.~Satoh, ``Side-channel attack standard evaluation board sasebo-w for smartcard testing,'' \emph{Power}, vol.~3, no. 2012, p. 400, 2012.

\bibitem{hettwer2020encoding}
B.~Hettwer, T.~Horn, S.~Gehrer, and T.~G{\"u}neysu, ``Encoding power traces as images for efficient side-channel analysis,'' in \emph{2020 IEEE International Symposium on Hardware Oriented Security and Trust (HOST)}.\hskip 1em plus 0.5em minus 0.4em\relax IEEE, 2020, pp. 46--56.

\bibitem{picek2023sok}
S.~Picek, G.~Perin, L.~Mariot, L.~Wu, and L.~Batina, ``Sok: Deep learning-based physical side-channel analysis,'' \emph{ACM Computing Surveys}, vol.~55, no.~11, pp. 1--35, 2023.

\bibitem{picek2018performance}
S.~Picek, I.~P. Samiotis, J.~Kim, A.~Heuser, S.~Bhasin, and A.~Legay, ``On the performance of convolutional neural networks for side-channel analysis,'' in \emph{Security, Privacy, and Applied Cryptography Engineering: 8th International Conference, SPACE 2018, Kanpur, India, December 15-19, 2018, Proceedings 8}.\hskip 1em plus 0.5em minus 0.4em\relax Springer, 2018, pp. 157--176.

\bibitem{kim2019make}
J.~Kim, S.~Picek, A.~Heuser, S.~Bhasin, and A.~Hanjalic, ``Make some noise. unleashing the power of convolutional neural networks for profiled side-channel analysis,'' \emph{IACR Transactions on Cryptographic Hardware and Embedded Systems}, pp. 148--179, 2019.

\bibitem{wu2020remove}
L.~Wu and S.~Picek, ``Remove some noise: On pre-processing of side-channel measurements with autoencoders,'' \emph{IACR Transactions on Cryptographic Hardware and Embedded Systems}, pp. 389--415, 2020.

\bibitem{kwon2021non}
D.~Kwon, H.~Kim, and S.~Hong, ``Non-profiled deep learning-based side-channel preprocessing with autoencoders,'' \emph{IEEE Access}, vol.~9, pp. 57\,692--57\,703, 2021.

\bibitem{tinyaes}
{K}okke, ``tiny-{AES}-c,'' \url{https://github.com/kokke/tiny-AES-c}, 2019.

\bibitem{hajra2013snr}
S.~Hajra and D.~Mukhopadhyay, ``Snr to success rate: reaching the limit of non-profiling dpa,'' \emph{Cryptology ePrint Archive}, 2013.

\bibitem{sha256}
{Intel}, ``{SHA256},'' \url{https://github.com/intel/tinycrypt}, 2017.

\bibitem{lz4}
{LZ4}, ``{LZ4},'' \url{https://github.com/lz4/lz4}, 2022.

\bibitem{ini}
``{inih},'' \url{https://github.com/benhoyt/inih}, 2021.

\bibitem{stm32CubeIDE}
``{Integrated Development Environment for STM32},'' \url{https://www.st.com/en/development-tools/stm32cubeide.html}, 2024.

\bibitem{zephyrSecureMqttSample}
``{Secure MQTT Sensor/Actuator},'' \url{https://github.com/zephyrproject-rtos/zephyr/tree/8a4b3a4b618a6b18390cfb4d67ad505abc1c83ab/samples/net/secure_mqtt_sensor_actuator}, 2024.

\bibitem{zephyr}
``{Zephyr Project},'' \url{https://www.zephyrproject.org/}, 2021.

\end{thebibliography}

\newpage
\section*{Appendix A}

\subsection{Experimental Setup}
\label{sec:appendix:exp_setup}
\begin{figure}[!h]
    \centering
    \includegraphics[width=5.8cm]{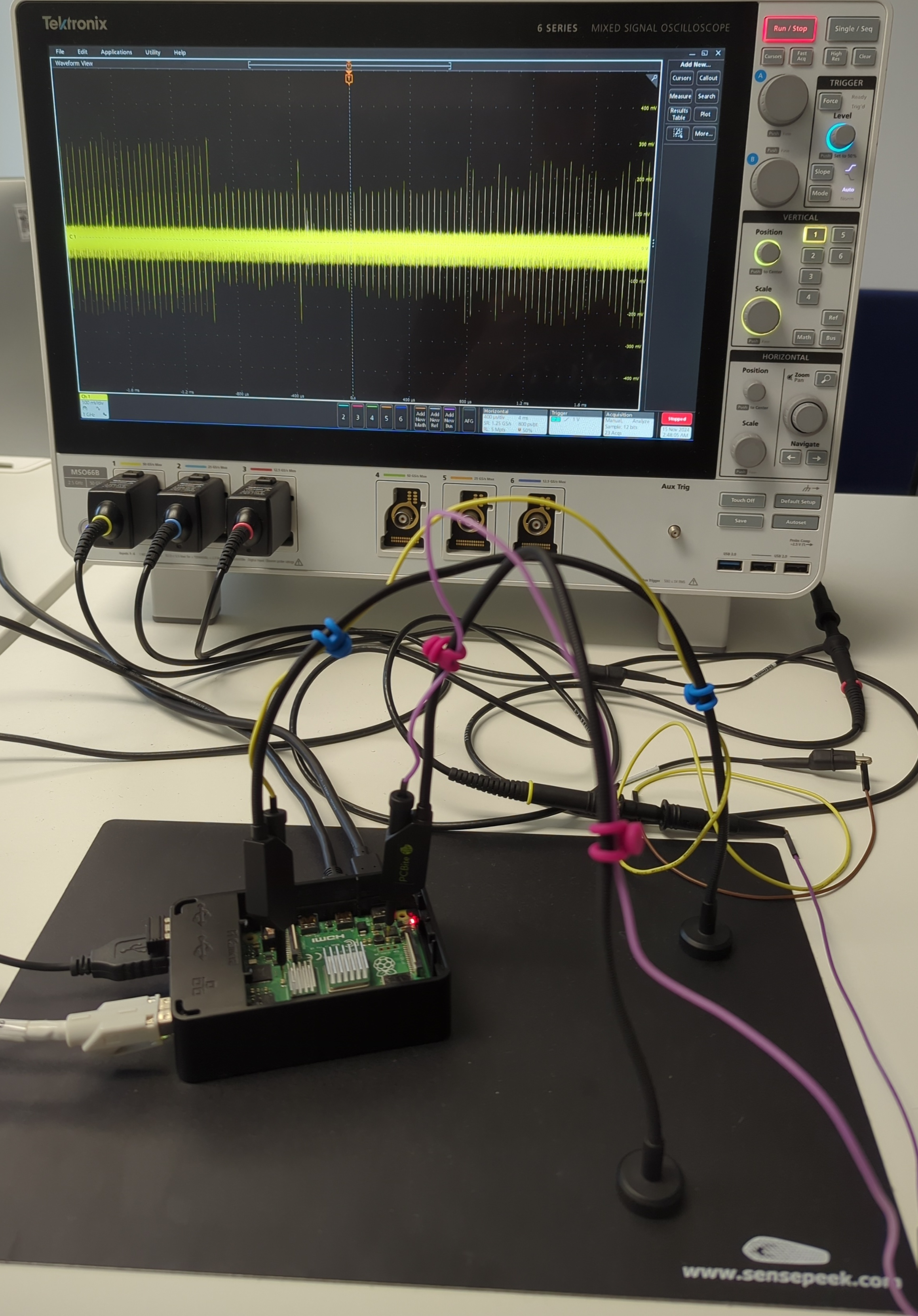}
    \caption{Our Experimental setup}
    \label{fig:Exp_Setup}
\end{figure}

Figure~\ref{fig:Exp_Setup} shows our experimental setup for the Raspberry Pi experiment. As mentioned before, we use a Tektronix MSO66B oscilloscope for power measurement.

\subsection{Confusion matrix for different probing points}
\label{sec:appendix:CM1}
\begin{figure}[!h]
    \centering
    \subfloat[\centering Probing on processor's pins (ACC=91\%)]{{\includegraphics[width=4.4cm]{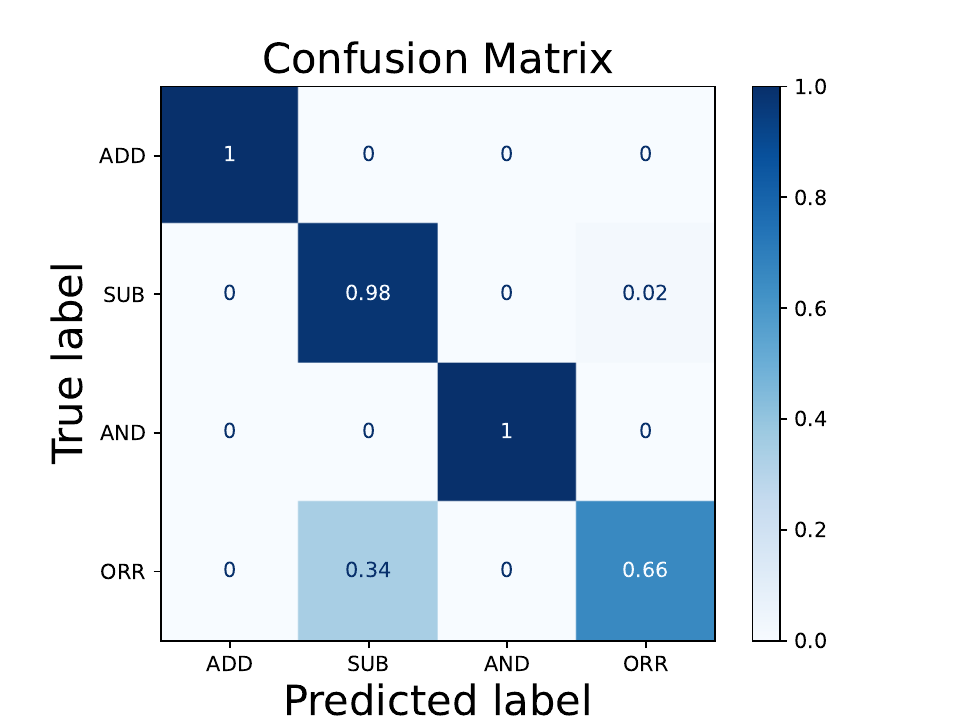}}}
    \subfloat[\centering Probing on board's pins (ACC=53\%)]{{\includegraphics[width=4.4cm]{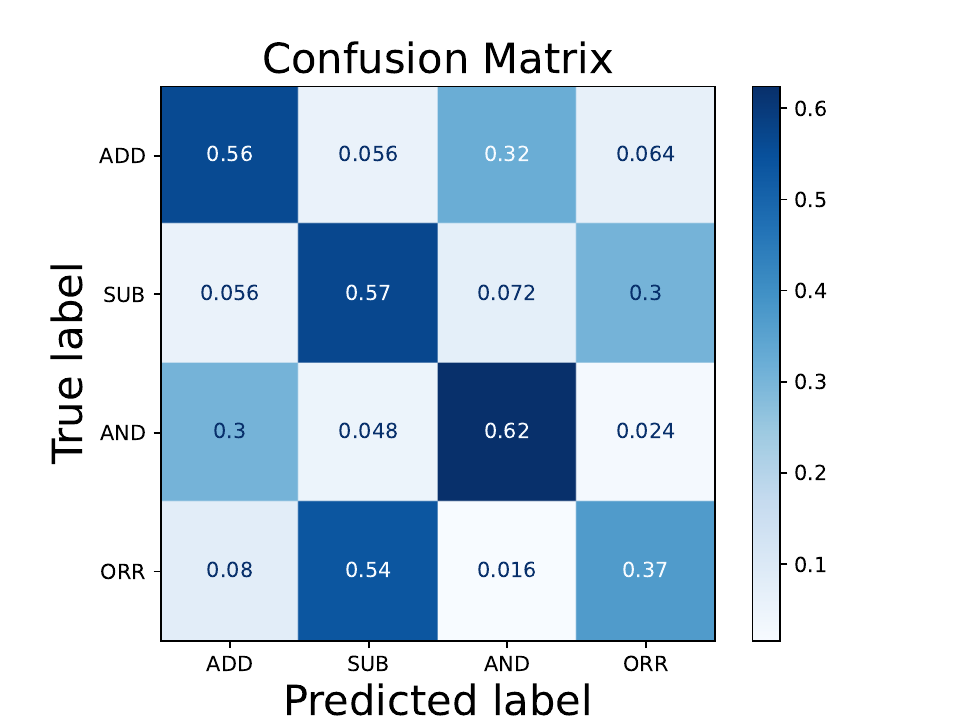}}}
    \caption{Classification accuracy for same firmware on the same target with different probing points using an MLP.}
    \label{fig:DirtyClean_Noise}
\end{figure}

Figure~\ref{fig:DirtyClean_Noise} shows the confusion matrix for two probing points. As illustrated in the figure, the accuracy drops drastically when the probing point is on the main power supply of the board.

\subsection{Confusion Matrix for Covariate Shift}
\label{sec:appendix:CM2}
\begin{figure}[!ht]
    \centering
    \subfloat[\centering Train and attack on data from the same distribution (Acc=81\%)]{
        \includegraphics[width=0.4\textwidth]{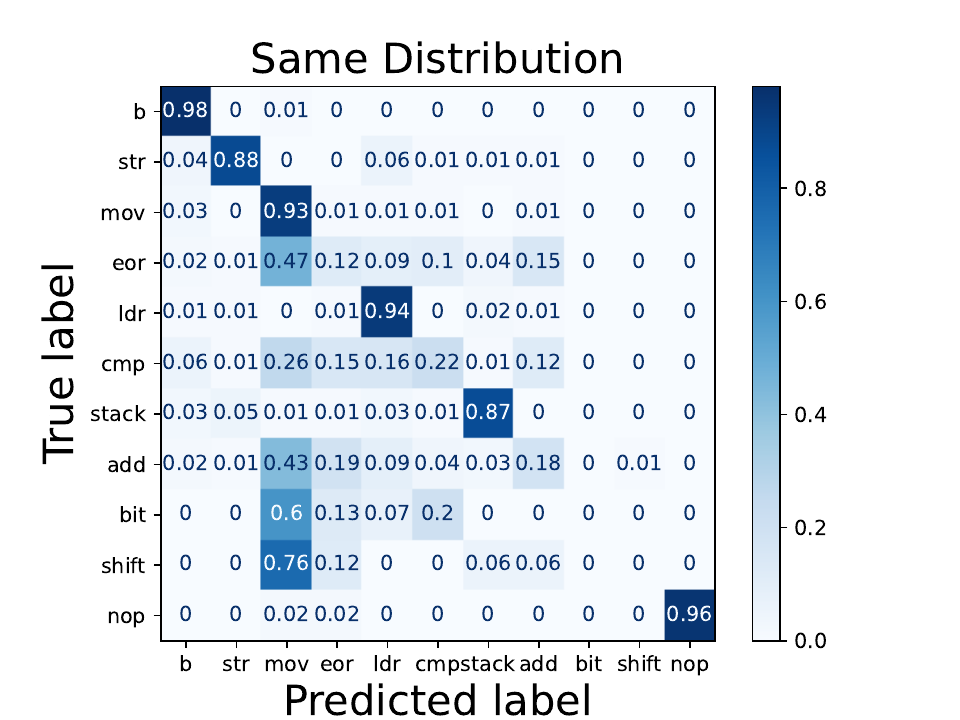}
    }
    \hfill 
    \subfloat[\centering Train on random data and attack on AES (Acc=41\%)]{
        \includegraphics[width=0.4\textwidth]{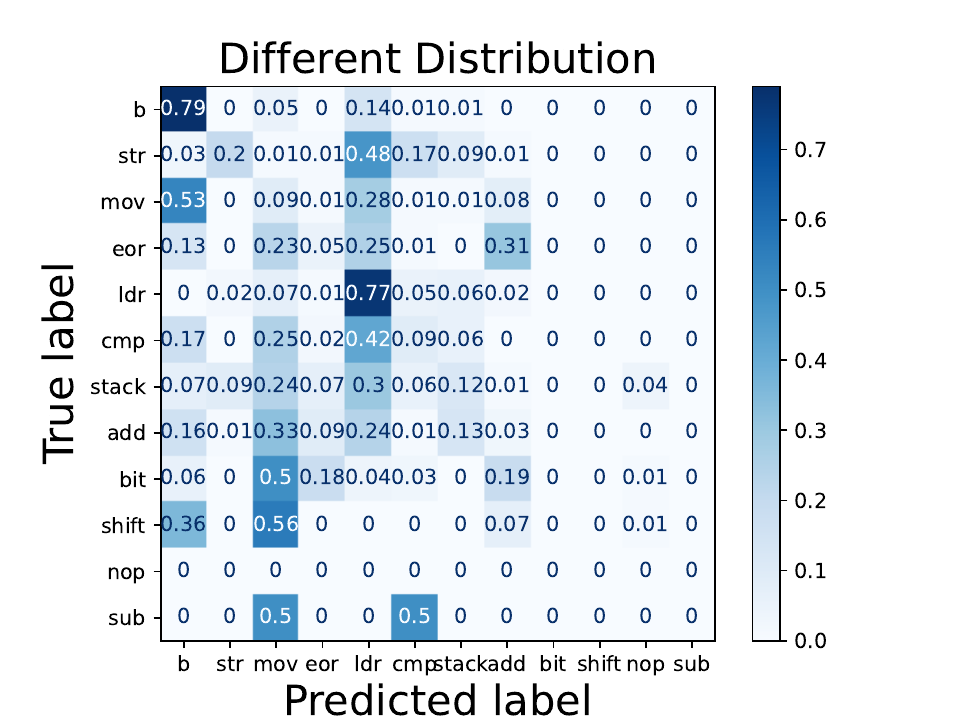}
    }
    \caption{Effect of covariate shift in instruction disassembly.}
    \label{fig:CovariateShift_CM}
\end{figure}

Figure~\ref{fig:CovariateShift_CM} shows the confusion matrix mentioned in the Section~\ref{sec:challenges:syssec:signalprocessing}.

\subsection{UBlox GNSS Receiver}
\label{sec:appendix:UBlox}
\begin{figure}[!h]
    \centering
    \subfloat[\centering With shield]{{\includegraphics[width=3.8cm]{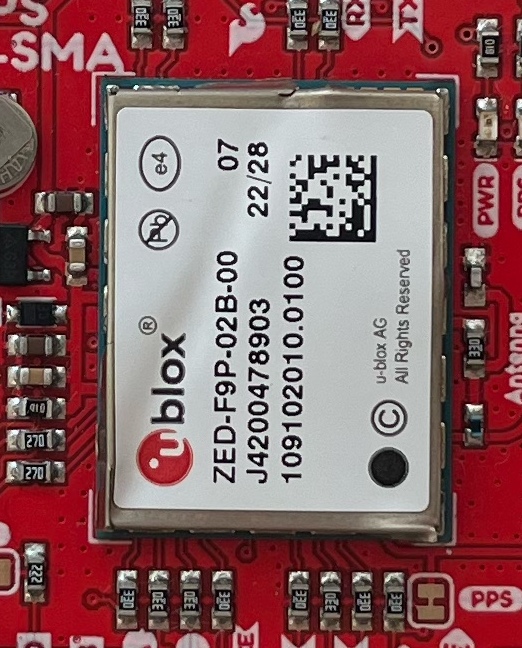}}}
    \qquad
    \subfloat[\centering Without shield]{{\includegraphics[width=3.8cm]{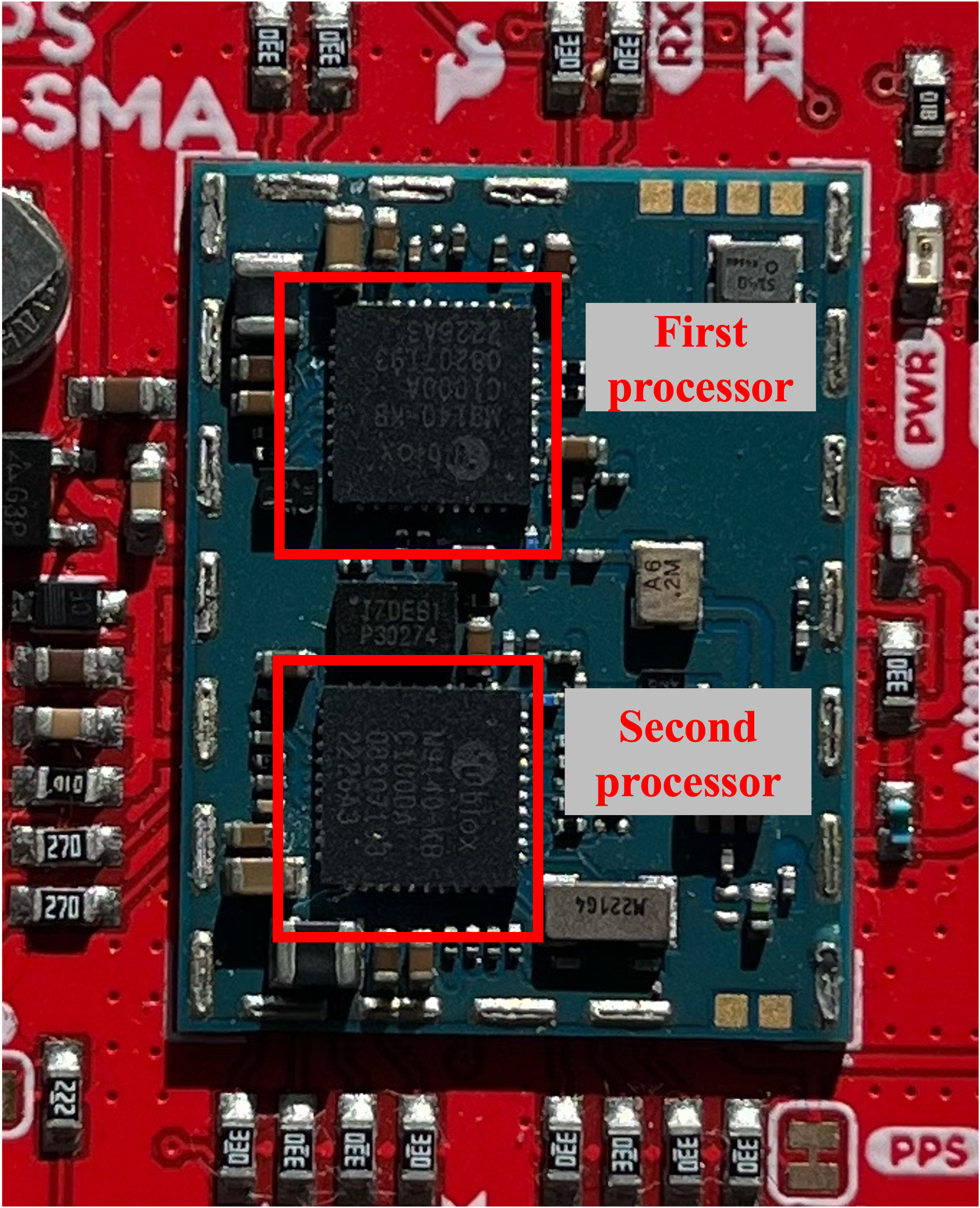}}}
    \caption{The UBlox ZED-F9P-02B GNSS receiver}
    \label{fig:UBlox_Unboxed}
\end{figure}

The UBlox GNSS receiver is shown Figure~\ref{fig:UBlox_Unboxed} in both with and without shield. In the unboxed figure (b), we see that there are two identical processors inside it (highlighted).


\end{document}